\documentclass[final]{siamltex}

\usepackage[dvips]{graphicx}
\usepackage{fancyhdr}
\usepackage{graphicx}
\usepackage{amssymb}

\title{Separatrix Map Analysis for Fractal Scatterings in Weak Interactions of Solitary Waves}

\author{Yi Zhu\thanks{Zhou Pei-Yuan Center for Applied Mathematics, Tsinghua University,
Beijing 100084, China({\tt zhuyi03@mails.tsinghua.edu.cn}).}
        \and Richard Haberman \thanks{Department of Mathematics, Southern Methodist University, Dallas, TX 75275,USA({\tt
        rhaberma@mail.smu.edu})}
        \and Jianke Yang \thanks{Department of Mathematics and Statistics, University of Vermont,
        16 Colchester Avenue, Burlington, VT 05401, USA({\tt
        jyang@cems.uvm.edu})}}
\begin{document}

\maketitle

\begin{abstract}
Previous studies have shown that fractal scatterings in weak
interactions of solitary waves in the generalized nonlinear
Schr\"odinger equations are described by a universal second-order
separatrix map. In this paper, this separatrix map is analyzed in
detail, and hence a complete characterization of fractal scatterings
in these weak interactions is obtained. In particular, scaling laws
of these fractals are derived analytically for different initial
conditions, and these laws are confirmed by direct numerical
simulations. In addition, an analytical criterion for the occurrence
of fractal scatterings is given explicitly.
\end{abstract}


%\begin{AMS}
%15A15, 15A09, 15A23
%\end{AMS}

\pagestyle{myheadings} \thispagestyle{plain} \markboth{Y.Zhu, R.
Haberman and J. Yang}{Separatrix map analysis for fractal
scatterings}

\section{Introduction}
Solitary wave interactions are important phenomena in science and
engineering \cite{Ablowitz_Segur,Hasegawa_Kodama}. These
interactions can be divided roughly into two types depending on the
strength of the interactions. Strong interactions, often called
collisions, are the interactions of solitary waves at close
distance. They would occur when two solitary waves are initially far
apart but move toward each other at moderate or large speeds. Weak
interactions are the interactions of solitary waves at far distance
through weak tail overlap. These interactions would occur if the two
waves are initially well separated, and their relative velocities
are small or zero. In integrable systems, collisions of solitary
waves are elastic \cite{Ablowitz_Segur}, and their weak interactions
exhibit interesting but still simple behaviors
\cite{Hasegawa_Kodama,Karpman_Solovev,Gerdjikov,Gerdjikov_Doktorov_Yang,Yang_Manakov}.
For certain integrable systems perturbed by higher-order
corrections, if they can be asymptotically transformed to integrable
equations, then their solitary wave interactions would closely
resemble those in integrable systems
\cite{Kodama,Fokas_Liu,Marchant}. If the systems are non-integrable,
however, solitary wave interactions can be extremely complicated,
and they can depend on initial conditions in a sensitive, fractal
manner. This fractal-scattering phenomenon was first discovered for
kink and antikink collisions in the $\phi^4$ model
\cite{Ablowitz_Kruskal,Campbell1,anninos}, later in several other
physical systems as well \cite{Kivshar1, YangTan}. For these strong
interactions, a resonant energy exchange mechanism between the
collision and internal/radiation modes was found responsible for
this fractal scattering. To analyze these fractal scatterings,
approximate collective-coordinate ODE models based on variational
methods \cite{Malomed} have been derived, and these ODEs are found
to exhibit qualitatively similar fractal scatterings as in the PDEs
\cite{anninos,Kivshar2,Ueda_Kath,TanYang}. Goodman and Haberman
further studied these collective-coordinate ODE models using
dynamical systems methods \cite{Goodman_Haberman1,
Goodman_Haberman2, Goodman_Haberman3, Goodman_Haberman4,Goodman}.
Performing asymptotic analysis along separatrix (homoclinic) orbits,
they derived separatrix maps which led to the prediction of
$n$-bounce resonance windows. It is noted that the separatrix maps
derived in \cite{Goodman_Haberman1, Goodman_Haberman2,
Goodman_Haberman3, Goodman_Haberman4,Goodman} contain parameters
which depend on initial conditions. In addition, these maps differ
from one PDE system to another. On weak interactions, fractal
scatterings have been found as well in a weakly discrete sine-Gordon
equation and a class of generalized nonlinear Schr\"odinger (NLS)
equations \cite{Dmitriev_Kivshar,Dmitriev,zhuyang}. For these weak
interactions, the mathematical analysis can be made more rigorous
and quantitative. Indeed, by extending the Karpman-Solovev
perturbation method \cite{Karpman_Solovev}, Zhu and Yang derived a
simple and asymptotically accurate ODE model for weak interactions
in the generalized NLS equations with arbitrary nonlinearities
\cite{zhuyang}. After various normalizations, this ODE system
contains only a single constant parameter which corresponds to
different nonlinearities in the PDEs. These ODEs are a
two-degrees-of-freedom Hamiltonian system with highly coupled
potentials, and their forms are quite different from the
collective-coordinate ODE models derived and studied previously for
strong interactions \cite{anninos,Kivshar2,Ueda_Kath,TanYang,
Goodman_Haberman1, Goodman_Haberman2, Goodman_Haberman3,
Goodman_Haberman4}. Zhu, Haberman and Yang further analyzed this ODE
model and derived a simple second-order map by using asymptotic
methods near separatrix orbits
\cite{zhuhabermanyang2,zhuhabermanyang}. A remarkable feature of
this map is that it does not contain any free parameters after
various rescalings, thus it is universal for all weak interactions
of solitary waves in the generalized NLS equations with arbitrary
nonlinearities. Despite its simplicity, this map can capture all the
fractal-scattering phenomena of the original PDEs and the reduced
ODEs very well both qualitatively and quantitatively
\cite{zhuhabermanyang2}. Reduction of weak-interaction dynamics from
the PDEs into a simple and universal second-order map is the main
contribution of \cite{zhuyang,zhuhabermanyang2,zhuhabermanyang}.
With the availability of this universal map, one may now expect a
complete characterization and understanding of fractal scatterings
in weak wave interactions. For instance, we now would like to know
under what conditions fractal scatterings would occur or would not
occur. For another instance, we now would like to know how these
fractals change as nonlinearities of the PDEs and initial conditions
of the solitary waves vary. In addition, we now would like to
understand how the zoomed-in structures of the fractal are related
to the original structures, and how these geometric structures
dictate the features of interaction dynamics. All these questions
can be answered by a careful analysis of this universal map.

In this paper, we analyze this universal map in detail, which will
provide a complete characterization of fractal scatterings in weak
wave interactions in the PDEs. First we will show that this map has
a fractal structure of its own. We will delineate the map's fractal
by tracking its singular curves. Then we will connect the map's
fractal to that of the PDE, and thus reach a deep understanding of
the PDE's fractal as well as its solution dynamics. In addition, we
will determine how the PDE's fractal changes when the soliton
parameters and the nonlinearity of the PDE vary. A precise
analytical criterion for the occurrence of fractal scatterings in
the PDEs will also be given explicitly. All our analytical results
are confirmed by direct numerical simulations. These results
significantly advance our understanding of fractal scatterings in
weak interactions of solitary waves.

\section{Previous work} \label{sec2}
First, we summarize previous relevant work which will form the basis
for our later analysis. We consider weak interactions in the
generalized NLS equations
\begin{equation}
iU_t+U_{xx}+N(|U|^2)U=0. \label{eqGNLS}
\end{equation}
These equations admit solitary waves of the form
\begin{equation}
U=\Phi(x-\xi)e^{i\phi},
\end{equation}
where $\Phi(\theta)$ is a localized positive function, $\xi=Vt+x_0$
is the wave's center position,
$\phi=V(x-\xi)/2+(\beta+V^2/4)t-\sigma_0$ is the phase function, and
$\beta$ is the propagation constant (which determines the amplitude
of the wave). This wave has four free parameters: velocity $V$,
amplitude parameter $\beta$, initial position $x_0$, and initial
phase constant $\sigma_0$. In weak interactions, two such solitary
waves are initially well separated with small relative velocities
and amplitude differences. Then they would interfere with each other
through tail overlapping. When time goes to infinity, they either
separate from each other at constant velocities, or form a bound
state. The exit velocity, defined as $\Delta
V_\infty=|V_2-V_1|_{t\to \infty}$, depends on the initial conditions
of the two waves. When the two waves form a bound state, we define
$\Delta V_\infty=0$. Throughout this paper, we label the left and
right waves by numbers 1 and 2 respectively.

In \cite{zhuyang}, we have shown that for a large class of
nonlinearities $N(|U|^2)$, this weak interaction depends on the
initial conditions in a sensitive, fractal manner. To analyze this
fractal-scattering phenomenon, an extended Karpman-Solov'ev
perturbation method was utilized, and the following simple set of
dynamical equations for soliton parameters were derived
\cite{zhuyang}:
\begin{eqnarray} \begin{array}{l}
\zeta_{\tau\tau}=\cos{\psi}e^{\zeta}, \\
%\psi_{\tau\tau}=h\sin{\psi}e^{\zeta}=(1+\varepsilon)\sin{\psi}e^{\zeta}
\psi_{\tau\tau}=(1+\varepsilon)\sin{\psi}e^{\zeta}.
\end{array}
\label{eqDyfinal}
\end{eqnarray}
Here
\begin{eqnarray} \label{scaling1}
\psi=\Delta\phi, \ \zeta=-\sqrt{\beta}\Delta\xi, \
\tau=\sqrt{\frac{16\beta^{3/2} c^2}{P}}\: t, \
\varepsilon=\frac{P}{2\beta P_\beta}-1,
\end{eqnarray}
$\Delta \xi$ and $\Delta \phi$ are the distance and phase difference
between the two waves, $\beta=(\beta_{1,0}+\beta_{2,0})/2$,
$\beta_{k,0} \; (k=1, 2)$ are the initial propagation constants of
the two waves, $c$ is the tail coefficient of the solitary wave with
propagation constant $\beta$, and $P(\beta)$ is the power function
of the wave. This ODE system is universal for the PDE
(\ref{eqGNLS}), and different nonlinearities $N(|U|^2)$ only
correspond to different constant parameter $\varepsilon$. If
$\varepsilon=0$ [such as when Eq. (\ref{eqGNLS}) is the original NLS
equation], the ODEs (\ref{eqDyfinal}) are integrable, and their
solutions have explicit functional expressions
\cite{zhuyang,zhuhabermanyang}. If $\varepsilon\ne 0$,
(\ref{eqDyfinal}) is not integrable. In this case, when $\varepsilon
> 0$ and under certain initial conditions, we have found that these
ODEs exhibit fractal scattering structures which agree with those in
the PDEs (\ref{eqGNLS}) both qualitatively and quantitatively. But
when $\varepsilon<0$, no fractal scatterings arise in these ODEs and
their corresponding PDEs under any initial conditions
\cite{zhuyang}.

In order to further understand these fractal scatterings, we have
analyzed the ODE system (\ref{eqDyfinal}) extensively in
\cite{zhuhabermanyang2,zhuhabermanyang} for $|\varepsilon| \ll 1$,
using perturbation methods near separatrix orbits. These ODEs are a
two-degree-of-freedom Hamiltonian system with the conserved
Hamiltonian
\begin{eqnarray}\label{Heps}
H(\zeta,\dot{\zeta},\psi,\dot{\psi})=E+\frac{\varepsilon}{2(1+\varepsilon)}\dot{\psi}^2,
\end{eqnarray}
where
\begin{eqnarray} \label{Edef}
E=\frac{1}{2}(\dot{\zeta}^2-\dot{\psi}^2)-e^\zeta\cos{\psi}
\end{eqnarray}
is called the energy. We also define the momentum $M$ of Eqs.
(\ref{eqDyfinal}) as
\begin{equation} \label{Mdef}
M = \dot{\zeta}\dot{\psi}-e^\zeta\sin{\psi}.
\end{equation}
Both $E$ and $M$ are conserved when $\varepsilon=0$, but vary over
time when $\varepsilon\ne 0$. If the orbits are escape orbits where
$\zeta_\infty \equiv \zeta|_{\tau\to \infty}= -\infty$ (i.e. the two
solitary waves eventually separate from each other after weak
interactions), then the exit velocities $|\dot{\zeta}|_\infty$ can
be calculated from Eqs. (\ref{Heps}) and (\ref{Mdef}) as
\begin{equation}\label{zetaM}
|\dot{\zeta}|_\infty=\sqrt{H+\sqrt{H^2+M^2_\infty/(1+\varepsilon)}}.
\end{equation}
In addition, $\dot{\psi}_\infty$, which determines the amplitudes of
exiting solitary waves, can also be obtained as
$\dot{\psi}_\infty=M_\infty/\dot{\zeta}_\infty$. Thus $M_\infty$ is
a key parameter for the prediction of weak-interaction outcomes. In
order to calculate $M_\infty$, we notice that for weak wave
interactions and when $|\varepsilon| \ll 1$, the ODE solutions [such
as $\zeta(\tau)$] oscillate near a sequence of separatrix orbits of
the unperturbed ($\varepsilon=0$) system before they escape to
infinity. On these separatrix orbits, $E=M=0$. If we consecutively
enumerate the minimums of $\zeta(\tau)$ (where interactions are the
weakest) and denote their energy and momentum values as $E_n$ and
$M_n$ (where $n$ is the index of the $\zeta$-minimum), then we can
analytically calculate $E_n$ and $M_n$ successively by integrating
along the separatrix orbits of the unperturbed system. This was done
in \cite{zhuhabermanyang2,zhuhabermanyang}, and we found that for
$|\varepsilon| \ll 1$, $E_n$ does not change, i.e. $E_n=E_0$ for all
$n\ge 1$. But $M_n$ does change. In the asymptotic limit of $E_n\ll
1, M_n\ll 1$ and $M_n/E_n\ll 1$, we found that the change of $M_n$
is asymptotically governed by the following second-order separatrix
map
\begin{eqnarray}
&&M_{n+1}=M_n-\mbox{sgn}(Q_n)\frac{8|E_0|^3\varepsilon}{\pi Q_n^2},  \label{map1}\\
&&Q_{n+1}=Q_n+2M_{n+1}, \label{map2}
\end{eqnarray}
with initial conditions $M_0$ and $Q_0$, where $Q_0=-S_0M_0$,
$S_0=S|_{\tau=0}$, and
\begin{equation}\label{S} S= \frac{2|C|^2\mbox{Im}(F)}{\pi
\mbox{Re}(C)}, \quad C = \sqrt{\frac{E+iM}{2}}, \quad F=
-\frac{1}{C}\mbox{acoth}(\frac{\dot{\zeta}+i\dot{\psi}}{2C}).
\end{equation}
Here the multi-valued functions $\sqrt{\cdot}$ and
$\mbox{acoth}(\cdot)$ are chosen uniquely by requiring
$\mbox{Im}(\sqrt{\cdot})\geq 0$ and
$\mbox{Im}(\mbox{acoth}(\cdot))\in [0, \pi)$ at the initial time.
The variable $Q_n$ in the above map is an auxiliary variable which
is related to the function $S$.

When the ODEs (\ref{eqDyfinal}) are integrable ($\varepsilon=0$),
$S$ is a conserved quantity. In this case, the integrable solution
$\zeta(\tau)$ develops finite-time singularity if $S_0=2k,
\mbox{Re}(C_0)\neq0, k=0,\pm1,\pm2,...$, or $\mbox{Re}(C_0)=0,
\mbox{Im}(F_0)=0$, see \cite{zhuhabermanyang} for details. These
finite-time singularities play important roles in the formation of
fractal scatterings. Indeed, it was observed from numerical
simulations in \cite{zhuyang} that fractal structures for
$\varepsilon\ne 0$ bifurcate out from points where integrable
solutions develop finite-time singularities. This fact will be
proved in this paper by the analysis of the map.

The map (\ref{map1})-(\ref{map2}) can be normalized into a very
simple form. Let
\begin{equation} \label{scaling}
G=\frac{8|E_0|^3\varepsilon}{\pi}, \quad m_n=G^{-1/3}M_n, \quad
q_n=G^{-1/3}Q_n,
\end{equation}
then this map becomes
\begin{eqnarray}
&&m_{n+1}=m_n-\frac{\mbox{sgn}(\varepsilon q_n)}{q_n^2}, \label{pmap1}\\
&&q_{n+1}=q_n+2m_{n+1}\label{pmap2}.
\end{eqnarray}
This normalized map is second-order, and it does not contain any
free parameters (except a sign of $\varepsilon$). Thus it is
universal for all weak two-wave interactions in the generalized NLS
equations (\ref{eqGNLS}) with arbitrary nonlinearities. For positive
and negative signs of $\varepsilon$, we will show this map has
completely different behaviors. This explains why fractal
scatterings appear in ODEs (\ref{eqDyfinal}) only when
$\varepsilon>0$, but not when $\varepsilon <0$. Details will be
given in Secs. \ref{seODE} and \ref{nepsi}.

To demonstrate the validity and accuracy of the above simple
separatrix map for describing fractal scatterings in the original
PDEs (\ref{eqGNLS}), here we compare the fractal scattering
structures obtained from the PDEs (\ref{eqGNLS}), the ODEs
(\ref{eqDyfinal}), and the map (\ref{pmap1})-(\ref{pmap2}). In all
our comparisons below and in later sections, we take the
cubic-quintic nonlinearity
\begin{equation}  \label{cqn}
N(|U|^2)=|U|^2+\delta |U|^4
\end{equation}
with $\delta=0.0003$ in the PDE (\ref{eqGNLS}) (comparisons with
other forms of nonlinearities are similar, see \cite{zhuyang}). We
take two types of initial conditions for the two solitary waves in
the PDE (\ref{eqGNLS}). One is that
\begin{equation} \label{ic1}
\beta_{0,1}=\beta_{0,2}=1,
\end{equation}
where the two waves initially have the same amplitudes. The other
one is that
\begin{equation} \label{ic2}
\beta_{0,1}=1.0325, \quad \beta_{0,2}=0.9675,
\end{equation}
where the two waves initially have unequal amplitudes. In both
cases, the other parameters in the two initial solitary waves are
the same as
\begin{equation} \label{icsame}
x_{0,2}=-x_{0,1}=5, \quad V_{0,1}=V_{0,2}=0, \quad \phi_{0,1}=0.
\end{equation}
That is, the two waves initially have equal velocities and are
separated by 10 spatial units. The initial phase of the first wave
$\phi_{0,1}$ can always be set as zero by phase invariance of the
PDE (\ref{eqGNLS}), thus it does not constitute a restriction on the
initial conditions. For both types of initial conditions,
$\Delta\phi_0 (=\phi_{0,2})$ is used as the control parameter.
Corresponding to both types of initial conditions, we find from Eq.
(\ref{scaling1}) that $\varepsilon=0.001$ in the ODEs
(\ref{eqDyfinal}).

For the first type of initial conditions (\ref{ic1}), the
corresponding initial conditions of the ODEs are
\begin{eqnarray}
\quad \quad \zeta_0=-10, \quad
\dot{\zeta}_0=\dot{\psi}_0=0,\label{ic1a}
\end{eqnarray}
and $\psi_0 (=\phi_{0,2})$ is the control parameter. For the map
(\ref{pmap1})-(\ref{pmap2}), the corresponding initial conditions
can be readily found to be
\begin{equation}\label{iv1}
q_0=m_0=-\frac{1}{2}\pi^{1/3}\tan\psi_0\cdot \varepsilon^{-1/3}
\end{equation}
since $S_0=-1$ in view of the expression (\ref{S}) for $S$. Here we
have assumed $\cos\phi_0>0$ since we have shown before
\cite{zhuhabermanyang} that fractal scatterings arise only in this
case.

For the second type of initial conditions (\ref{ic2}), the
corresponding initial conditions of the ODEs are
\begin{equation}
\quad \quad \zeta_0=-10, \quad \dot{\zeta}_0=0, \quad
\dot{\psi}_0=-0.01167. \label{ic1b}
\end{equation}
The corresponding initial conditions of the map are
\begin{eqnarray}
m_0=T(\psi_0)\varepsilon^{-1/3}, \qquad
q_0=-T(\psi_0)S(\psi_0)\varepsilon^{-1/3},   \label{iv2}
\end{eqnarray}
where
\begin{equation}\label{Tdef}
T(\psi_0)=\frac{-\pi^{1/3}
\sin{\psi_0}}{2\cos{\psi_0}+\dot{\psi}_0^2e^{-\zeta_0}},
\end{equation}
and $S(\psi_0)$ is defined in (\ref{S}).

\begin{figure}
\center
\includegraphics[width=100mm,height=80mm]{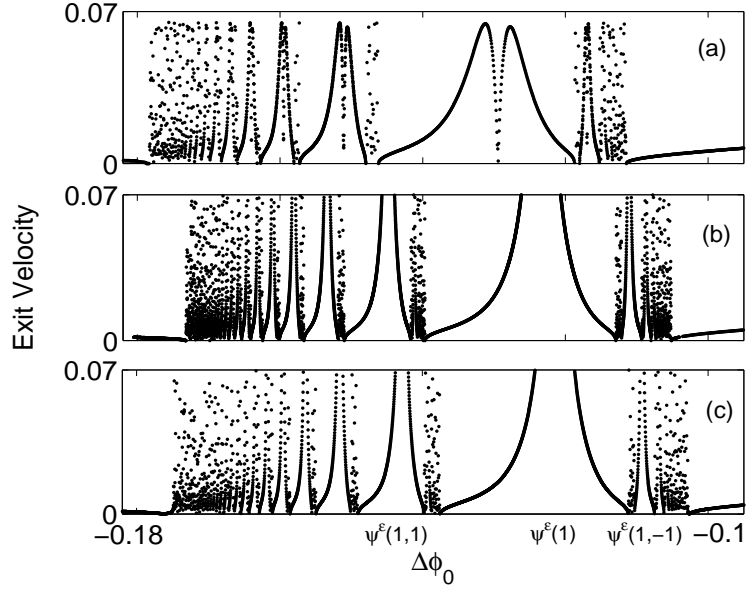}
\caption{\label{comparison} Comparisons of exit-velocity fractals
from (a) PDE simulations, (b) ODE predictions, and (c) map
predictions for the equal-amplitude initial conditions (\ref{ic1}).
The marked segment in (b) is amplified in Fig. \ref{dynamics}, where
the solution dynamics in this segment is also shown. Labels
$\psi^\varepsilon(1), \psi^\varepsilon(1, 1), \dots$ are locations
of singularity peaks (of infinite height) in (c), which correspond
to intersections of the initial-value curve $\lambda^\varepsilon$
with the map's singular curves $\gamma(1), \gamma(1,1), \dots$ in
Fig. \ref{iccurve1} (see Sec. \ref{sec_ic1} for details). }
\end{figure}

\begin{figure}
\center
\includegraphics[width=100mm,height=80mm]{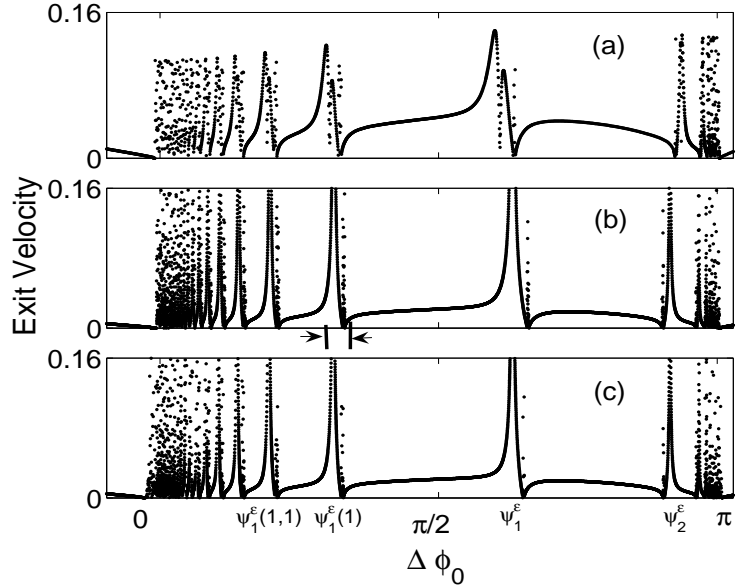}
\caption{\label{comparison2} Comparisons of exit-velocity fractals
from (a) PDE simulations, (b) ODE predictions, and (c) map
predictions for the unequal-amplitude initial conditions
(\ref{ic2}). For the segment of structures marked in (b), its
movement with varying values of $\varepsilon$ is displayed in Fig.
\ref{un2zero}. Labels $\psi_1^\varepsilon, \psi_1^\varepsilon(1),
\dots$ are locations of singularity peaks in (c) which correspond to
intersections of the initial-value curve $\lambda^\varepsilon$ with
the map's singular curves $\gamma_0, \gamma(1), \dots$ in Fig.
\ref{iccurve2} (see Sec. \ref{sec_ic2} for details).  }
\end{figure}

For each of the above two types of initial conditions, we have
simulated the PDE (\ref{eqGNLS}), the ODEs (\ref{eqDyfinal}), and
the map (\ref{pmap1})-(\ref{pmap2}) for various values of the
control parameter $\Delta\phi_0$, and obtained the corresponding
exit velocities $\Delta V_\infty$ [in the case of the ODEs and the
map, the exit velocities have been obtained and properly rescaled
through the formula (\ref{zetaM}) and relations (\ref{scaling1}),
(\ref{scaling})]. In all cases, we observed fractal scattering
phenomena. For the first type of initial conditions (\ref{ic1})
(where the two waves initially have equal amplitudes), the fractal
structures from the PDE, ODE and the map are displayed in Fig.
\ref{comparison}. Here the fractals are symmetric in $\Delta\phi_0$,
and they lie in a narrow interval near the $\Delta\phi_0=0$ point
(only a segment of the parameter region is shown for better
visualization). We see that the three structures match very well
both qualitatively and quantitatively. For the second type of
initial conditions (\ref{ic2}) (where the two waves initially have
un-equal amplitudes), the fractal structures from the PDE, ODE and
the map are displayed in Fig.\ref{comparison2}. Here the fractals
lie in a wide interval of $\Delta\phi_0\in [0, \pi]$, and they do
not possess any symmetry. Again these structures match very well.
These comparisons show that the intricate fractal scatterings in the
PDEs (\ref{eqGNLS}) and the ODEs (\ref{eqDyfinal}) CAN be accurately
predicted by the simple map (\ref{pmap1})-(\ref{pmap2}). These are
the main results which have been obtained before in
\cite{zhuyang,zhuhabermanyang2,zhuhabermanyang}.

In order to reach a complete understanding and characterization of
these fractal structures and their solution dynamics in weak wave
interactions, the separatrix map (\ref{pmap1})-(\ref{pmap2}) needs
to be carefully analyzed. This map exhibits a fractal structure of
its own in its initial-condition space (when
$\mbox{sgn}(\varepsilon)=1$) \cite{zhuhabermanyang2}. If this
fractal of the map is well understood, then through connections of
variables between the map and the PDEs/ODEs, the fractals in the
PDEs/ODEs will be well characterized. In the following sections, we
will analyze this map and use this information to delineate
weak-interaction dynamics in the PDEs/ODEs.

\section{Analysis of the fractal in the separatrix map}
\label{sec_3}

The separatrix map (\ref{pmap1})-(\ref{pmap2}) depends only on the
sign of $\varepsilon$, which can be 1 or $-1$. It turns out that
this map has completely different dynamics for
$\mbox{sgn}(\varepsilon)=1$ and $-1$. When
$\mbox{sgn}(\varepsilon)=1$, this map is fractal-bearing, while when
$\mbox{sgn}(\varepsilon)=-1$, it is not \cite{zhuhabermanyang2}.
This is consistent with the numerical observation of \cite{zhuyang}
that fractal scatterings occur in the ODEs (\ref{eqDyfinal}) only
when $\varepsilon>0$ but not when $\varepsilon<0$. In this section,
we analyze the fractal in this map when $\mbox{sgn}(\varepsilon)=1$.

For convenience, we rewrite the map (\ref{pmap1})-(\ref{pmap2}) as
\begin{eqnarray}
\left(\begin{array}{l}q_{n+1}\\m_{n+1}\end{array}\right)=\mathcal{F}\left(\begin{array}{l}q_n\\m_n\end{array}\right),
\end{eqnarray}
where
\begin{eqnarray}\label{F}
\mathcal{F}\left(\begin{array}{l}q\\m\end{array}\right)=\left(\begin{array}{l}q+2m-\frac{2
sgn(q)}{q^2}\\m-\frac{{sgn}(q)}{q^2}\end{array}\right).
\end{eqnarray}
It can be seen that $\mathcal{F}$ is differentiable when $q\ne 0$.
The Jacobian of $\mathcal{F}$ is
\begin{eqnarray}
J_{\mathcal{F}}=\left(\begin{array}{cc}1+4\frac{{sgn}(q)}{q^3}\quad
&2\\2\frac{{sgn}(q)}{q^3} &1\end{array}\right).
\end{eqnarray}
The determinant of this Jacobian matrix is equal to one, thus the
map $\mathcal{F}$ is area-preserving and orientation-preserving.
Chaos if it occurs will be Hamiltonian chaos \cite{Arrowsmith}.

It is easy to see that
$\mathcal{F}^{n}(-q,-m)=-\mathcal{F}^{n}(q,m)$, thus $\mathcal{F}^n$
is anti-symmetric with respect to the origin. It can also be seen
that $\mathcal{F}$ does not have any (bounded) fixed points, but it
has a lot of periodic orbits. For example, \{$(2^{-1/3},2^{-1/3})$,
$(-2^{-1/3},-2^{-1/3})$\} is a period-2 orbit, and
\{$(1,1),(1,0),(-1,-1),(-1,0)$\} is a period-4 orbit. One can easily
verify that these two periodic orbits are both unstable. All the
other periodic orbits are unstable too as we will see later in this
section.

An important property of the map $\mathcal{F}$ is that it is
reversible in $\mathbb{R}^2\backslash\Omega$, where $\mathbb{R}$ is
the set of real numbers, and
\[\Omega=\{(q,m)\in \mathbb{R}^2: q=2m\}.\]
Indeed, $\mathcal{F}^{-1}$ has an explicit formula when
$q\ne 2m$:
\begin{eqnarray} \label{Finv}
\mathcal{F}^{-1}\left(\begin{array}{l}q\\
m\end{array}\right)=
\left(\begin{array}{l}q-2m\\m+\frac{{sgn}(q-2m)}{(q-2m)^2}\end{array}\right).
\end{eqnarray}
In the later text, we will utilize two subsets of the $(q, m)$ plane
which we define here:
\[\Omega^+=\{(q,m)\in \mathbb{R}^2: q-2m>0\}, \]
and
\[\Omega^-=\{(q,m)\in \mathbb{R}^2: q-2m<0\}.\]

\subsection{Singular curves of the map and their
identifications}

\begin{figure}
\includegraphics[width=130mm,height=100mm]{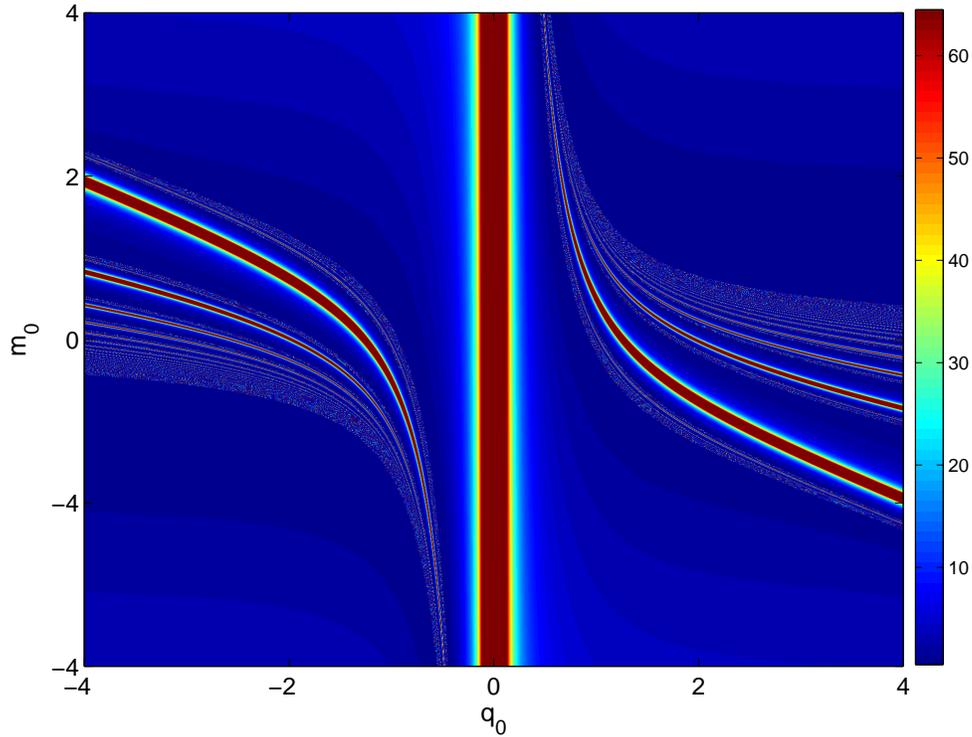}
\caption{\label{2dmap} Values of $|m_\infty|$ in the initial-value
space $(q_0, m_0)$ for the map (\ref{pmap1})-(\ref{pmap2}) with $
\mbox{sgn}(\varepsilon)=1$. Colors represent value levels of
$|m_\infty|$. }
\end{figure}

The limit behaviors $\lim_{n\to \infty}\mathcal{F}^n(q_0,m_0)$ are
important as they correspond to the outcomes of weak interactions.
Indeed, from $m_\infty=\lim_{n\to \infty}m_n$ and the scaling
relations (\ref{scaling}), we can obtain the exit velocities
$|\dot{\zeta}_\infty|$ from Eq. (\ref{zetaM}). For almost all
initial points $(q_0,m_0)$, $m_\infty$ exists and is finite (while
$q_\infty=\pm \infty$). Thus orbits of almost all initial points
eventually escape to $q_\infty=\infty$ or $-\infty$ along the
horizontal (constant-$m$) direction. But there are two special sets
of initial points which are different. One set is the initial points
which make $q_n=0$ for some $n\ge 0$, i.e.
\begin{equation} \label{calSdef}
\mathcal{S}=\{(q_0,m_0)\in \mathbb{R}^2: q_n=0,\, n=0,1,2,...\}.
\end{equation}
For these points, iterations can not continue for $(q_{n+1},
m_{n+1})$ since $q_n=0$, thus we call them singular points, and
$\mathcal{S}$ the singular set. On these points, we formally let
$m_{n+1}=\infty$, and consequently $m_\infty=\infty$ as well. For
singular initial points, the exit velocities of weak interactions
are infinite [see Eq. (\ref{zetaM})], thus they correspond to peaks
(of infinite height) in Fig. \ref{comparison}(c). The other set is
the initial points $(q_0,m_0)$ which are periodic or quasi-periodic.
On these points, $m_n$ oscillates forever, hence $m_\infty$ does not
exist. These points correspond to spatially-localized and temporally
oscillating bound states in the PDEs (\ref{eqGNLS}). In this case,
we set $m_\infty=0$, which gives zero separation velocities
$|\dot{\zeta}_\infty|$ from Eq. (\ref{zetaM}) (recall that $H<0$ for
fractal scatterings \cite{zhuhabermanyang}). This way, $m_\infty$
can be defined everywhere in the initial value plane $(q_0,m_0)$.
Numerically we have computed $|m_\infty|$ over this plane (iterating
500 steps instead of infinite steps), and the result is plotted in
Fig. \ref{2dmap} (here color levels correspond to $|m_\infty|$
values). This graph is a fractal, as can be easily verified by
repeated zooms into it.

In Fig. \ref{2dmap}, singular points form an infinite number of
smooth curves.  Each of these singular curves lies in the middle of
a red stripe (of varying thickness), and all these singular curves
form the backbone of the map's fractal in Fig. \ref{2dmap}. These
singular curves are directly related to the singularity peaks (of
infinite height) in the map's exit-velocity fractals in Figs.
\ref{comparison}(c) and \ref{comparison2}(c). This is because on a
singular curve, $m_\infty=\infty$, thus the corresponding exit
velocity is also infinite [see Eq. (\ref{zetaM})]. In view of this,
these singular curves correspond to the singularity peaks in the
ODE's exit-velocity fractals and counterpart structures in the PDE's
exit-velocity fractals. These singularity peaks (or their
counterparts) in turn form the backbones of the map's, ODE's and
PDE's exit-velocity fractals in Figs. \ref{comparison} and
\ref{comparison2}. Thus if we can clearly characterize the singular
curves of the map, then a good understanding of the fractals in the
ODEs and PDEs will be reached. In the rest of this section, we focus
on these singular curves. We will determine where they are located,
how to identify them, what dynamics they represent, and what their
asymptotics are at large or small values of $q_0$.

\begin{figure}
\includegraphics[width=130mm,height=100mm]{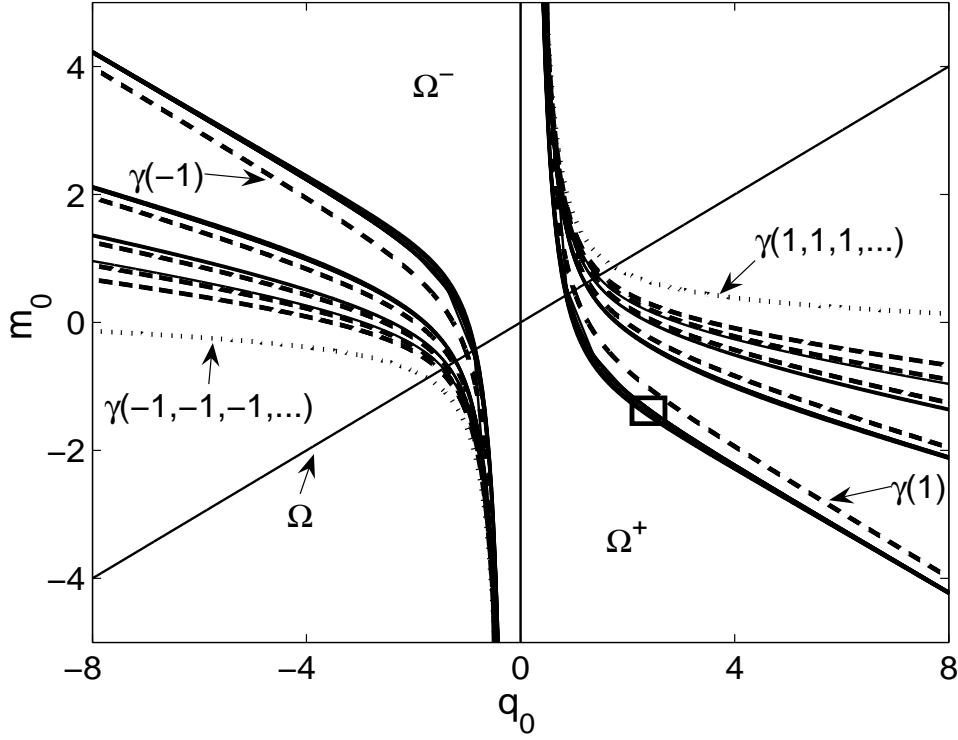}
\caption{Plots of singular curves $\gamma(a_0), \gamma(a_0, a_1),
\gamma(a_0, a_1, a_2)$, $\gamma(a_0, a_1, a_2, a_3)$, and
$\gamma(a_0, a_1, a_2, a_3, a_4)$ for all binary numbers of $a_0,
\dots, a_4$. The dashed lines in the right plane are $\gamma(1)$,
$\gamma(1,1)$, $\gamma(1,1,1)$, $\gamma(1,1,1,1)$ and
$\gamma(1,1,1,1,1)$ respectively from left to right. The
accumulation curve $\gamma(1,1,1, \dots)$ is also shown (as dotted
lines). Lines in the left half plane are similar. The box region
will be magnified in Fig.\ref{singularzoom}.  } \label{singular}
\end{figure}

About these singular curves, each one is characterized by a unique
finite binary sequence $\mathbf{a}= \mbox{sgn}(q_0,q_1,...,q_{n})$.
Singular points on the same curve have the same binary sequence,
while different singular curves have different sequences. Let us
denote the singular curve with a binary sequence
$\mathbf{a}=(a_0,a_1,...,a_{n})$ as $\gamma(\mathbf{a})$. Then
\begin{equation}  \label{gammaa}
\gamma(\mathbf{a})=\{(q_0,m_0)\in\mathbb{R}^2: \; q_{n+1}=0,\;
\mbox{sgn}(q_0,q_1,...,q_{n})=\mathbf{a}\}.
\end{equation}
Here we allow $\mathbf{a}$ to be empty, in which case we get the
simplest singular curve
\begin{equation}\label{gamma0}
\gamma_0=\{(q_0,m_0):q_0=0\},
\end{equation}
which is the vertical axis. When $n=0$ (where $q_1=0$), we see from
the map (\ref{F}) that
\begin{eqnarray}
&&\gamma(1)=\{(q_0,m_0): m_0=-q_0/2+1/q_0^2, \; q_0>0\},  \label{gamma1} \\
&&\gamma(-1)=\{(q_0,m_0): m_0=-q_0/2-1/q_0^2, \; q_0<0\}.
\label{gamman1}
\end{eqnarray}
These two curves are plotted in Fig. \ref{singular}. They are
located in the middle of the thickest red stripes in the right and
left half planes of Fig. \ref{2dmap}.

\begin{figure}
\includegraphics[width=130mm,height=100mm]{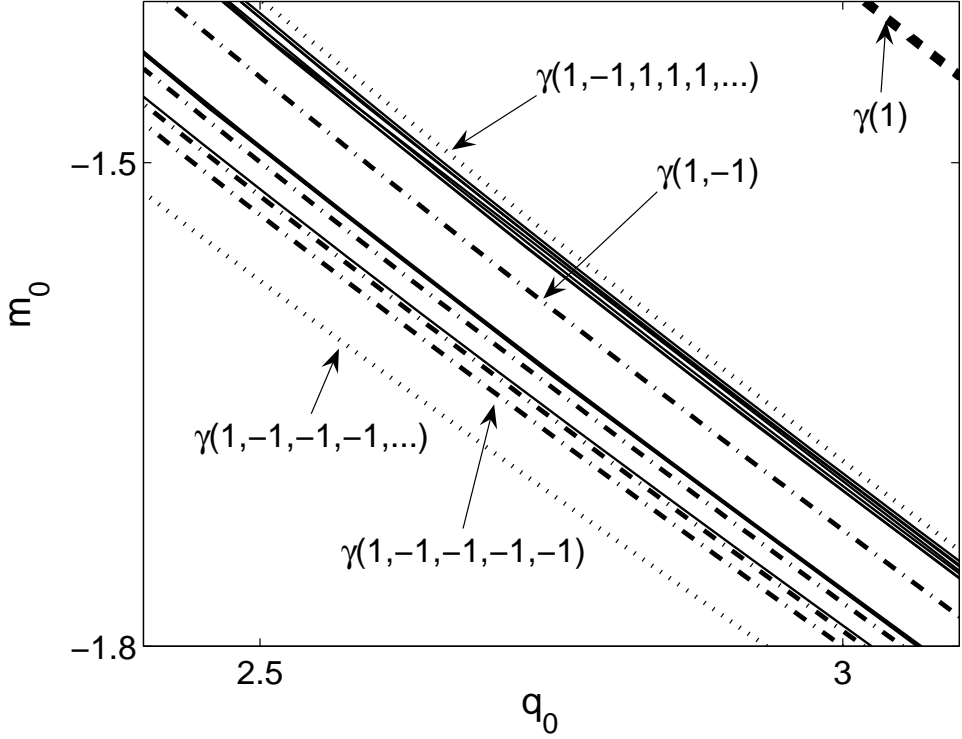}
\caption{\textit{Amplification of the box region in Fig.
\ref{singular}. The dash-doted lines are $\gamma(1,-1)$,
$\gamma(1,-1,-1)$, $\gamma(1,-1,-1,-1)$ and $\gamma(1,-1,-1,-1,-1)$
respectively from right to left. Accumulation curves $\gamma(1, -1,
-1, -1, \dots)$ and $\gamma(1, -1, 1, 1, 1, \dots)$ are also shown.
}} \label{singularzoom}
\end{figure}

Now we determine the relations between singular curves with
different binary sequences. From our definitions, we see that if
$(q_0, m_0) \in \gamma(a_0,\mathbf{a})$, then $\mathcal{F}(q_0,
m_0)\in \gamma(\mathbf{a})$, where $\mathbf{a}$ is any finite binary
sequence and $a_0=\pm1$. In addition, if $a_0=1$, i.e. $q_0>0$, then
$q_1-2m_1>0$ in view of Eq. (\ref{pmap2}), thus $\mathcal{F}(q_0,
m_0)\in \Omega^+$. Similarly if $a_0=-1$, then $\mathcal{F}(q_0,
m_0)\in \Omega^-$. As a result, we have
\begin{eqnarray}
&&\mathcal{F}(\gamma(1,\mathbf{a}))=\gamma(\mathbf{a})\bigcap\Omega^+,\\
&&\mathcal{F}(\gamma(-1,\mathbf{a}))=\gamma(\mathbf{a})\bigcap\Omega^-.
\end{eqnarray}
Or written differently, we have
\begin{eqnarray}
&&\gamma(1,\mathbf{a})=\mathcal{F}^{-1}(\gamma(\mathbf{a})\bigcap\Omega^+),\label{iFiterate1}\\
&&\gamma(-1,\mathbf{a})=\mathcal{F}^{-1}(\gamma(\mathbf{a})\bigcap\Omega^-).\label{iFiterate2}
\end{eqnarray}
These relations tell us that each singular curve
$\gamma(\mathbf{a})$ has two pre-image singular curves
$\gamma(1,\mathbf{a})$ and $\gamma(-1,\mathbf{a})$ under the map
$\mathcal{F}$. In particular, $\gamma(1)$ and $\gamma(-1)$ are the
two pre-image singular curves of $\gamma_0$.

Thus from the basic singular curve $\gamma_0$ given in Eq.
(\ref{gamma0}), two singular curves $\gamma(1)$ and $\gamma(-1)$ in
Eqs. (\ref{gamma1})-(\ref{gamman1}) are obtained. We can
successively construct $\gamma(\mathbf{a})$ for any finite binary
sequence $\mathbf{a}$ by repeatedly applying the inverse map
$\mathcal{F}^{-1}$ on $\gamma_0$. The first few of these singular
curves are plotted in Fig. \ref{singular}. It is easy to see from
the map (\ref{F}) that for any binary sequence $\mathbf{a}$,
$\gamma(\mathbf{a})$ and $\gamma(-\mathbf{a})$ are anti-symmetric to
each other about the origin, i.e.
\begin{equation} \label{gamma_symmetry}
\gamma(-\mathbf{a})=-\gamma(\mathbf{a}),
\end{equation}
thus we only consider curves on the right half plane with
$\mathbf{a}=(1,a_1,...,a_n)$ below. From Fig. \ref{singular}, we see
that $\gamma(1)$, $\gamma(1,1)$, $\gamma(1,1,1)$, ... form the
primary cascading sequence from the left to the right. This sequence
accumulates to the limit curve $\gamma(1,1,1,1,...)$ which is
plotted as a dotted line in Fig. \ref{singular} and can be seen in
Fig. \ref{2dmap}. This primary sequence of singular curves defines
the overall geometry of the map's fractal in Fig. \ref{2dmap}. On
the left side of each primary curve, there is a secondary structure
whose infinite curves are very close to each other and thus visually
show as one ``thick curve" in Fig. \ref{singular}. Each secondary
structure lies near its corresponding primary curve. To probe these
secondary structures, we zoom into the box region of Fig.
\ref{singular}, which contains a segment of the secondary structure
for the primary curve $\gamma(1)$. The result is shown in Fig.
\ref{singularzoom}. From this zoomed-in graph, we see that among
numerous curves in this secondary structure, there is a secondary
sequence of curves $\gamma(1,-1)$, $\gamma(1,-1,-1)$,
$\gamma(1,-1,-1,-1), \dots$ which cascades to the left. This
secondary curve sequence defines the overall geometry of this
secondary structure. Near each curve in this secondary sequence (and
on its right hand side), there is a higher-order structure which can
be probed by repeated zooms. Every time one zooms into a
higher-order structure, the cascading direction of its higher-order
sequence is reversed, and the side of the higher-order sequence
relative to its associated curve (either left or right) is also
reversed. The binary sequences for these higher-order-sequence
curves are \{$(\mathbf{a},1)$, $(\mathbf{a},1,1)$,
$(\mathbf{a},1,1,1)$, ...\} when the sequence cascades to the right
and lies on the right side of $\gamma(\mathbf{a})$, and
\{$(\mathbf{a}, -1)$, $(\mathbf{a}, -1, -1)$, $(\mathbf{a}, -1, -1,
-1)$ ...\} when the sequence cascades to the left and lies on the
left side of $\gamma(\mathbf{a})$. Here $\mathbf{a}$ is the binary
sequence for the associated curve of this higher-order sequence.

Based on the above pattern, we can identify any singular curve with
an arbitrary binary sequence. For instance, to identify the curve
with a binary sequence $\mathbf{a}=(1, 1, -1, -1, -1, 1, -1)$, we
first go to the curve $\gamma(1)$ (see Fig. \ref{singular}), find
its associated primary sequence (which lies on its right hand side),
and pick out the first member of that sequence (not counting
$\gamma(1)$ itself). The picked curve is then $\gamma(1,1)$, see
Fig. \ref{singular}.  Next, we go to the secondary curve sequence of
$\gamma(1,1)$ (which lies on its left) and pick out the third member
of that sequence, which is $\gamma(1,1, -1, -1, -1)$. Next we go to
the higher-order sequence of $\gamma(1,1, -1, -1, -1)$ (which lies
on its right) and pick out the first member of that sequence, which
is $\gamma(1,1, -1, -1, -1, 1)$. Lastly we go to the higher-order
sequence of $\gamma(1,1, -1, -1, -1, 1)$ (which lies on its left)
and pick out the first member of that sequence, which will be the
singular curve $\gamma(1,1, -1, -1, -1, 1, -1)$ that we are looking
for.

We have noted earlier in this section that the map $\mathcal{F}$ has
a lot of periodic orbits. Then an interesting question is where
these periodic orbits are located in the map's fractal in Fig.
\ref{2dmap}, and how these orbits are related to the above singular
curves. Clearly every periodic point can not lie on a singular curve
$\gamma(\mathbf{a})$ with a finite binary sequence $\mathbf{a}$ in
view of the definition (\ref{gammaa}) of $\gamma(\mathbf{a})$. But a
periodic point can be viewed as being on a singular curve with an
infinite binary sequence $\mathbf{a}$ whose digits are the signs of
$q$ of the successive iteration points which repeat with the same
period as the periodic point (here a singular curve with an infinite
binary sequence can be defined as the limit of the singular curve
with a finite binary sequence). Using this viewpoint, we can
understand where periodic points should be located in the map's
fractal. For instance, in the period-two orbit
\{$(2^{-1/3},2^{-1/3})$, $(-2^{-1/3},-2^{-1/3})$\}, the first point
$(2^{-1/3},2^{-1/3})$ lies on the singular curve with the binary
sequence $\mathbf{a}=(1, -1, 1, -1, \dots)$, while the second point
$(-2^{-1/3},-2^{-1/3})$ lies on the singular curve with the binary
sequence $\mathbf{a}=(-1, 1, -1, 1, \dots)$. For another instance,
in the period-four orbit \{$(1,1),(1,0),(-1,-1),(-1,0)$\}, the point
$(1,1)$ lies on the singular curve with a binary sequence $(1, 1,
-1, -1, 1, 1, -1, -1, \dots)$, and the point $(1, 0)$ lies on the
singular curve with a binary sequence $(1, -1, -1, 1, 1, -1, -1, 1,
\dots)$, etc. The singular curves with these infinite binary
sequences can be identified by the same scheme as we detailed above,
thus we can ascertain where these periodic points are located in the
map's fractal. Obviously these infinite binary sequences of periodic
points are infinitely close to their finite truncations, and points
on the singular curves with truncated finite binary sequences have
very different trajectories from those of the periodic points. Thus
the periodic points of the map $\mathcal{F}$ are all unstable.

From the above singular-curve identification scheme, we see that all
the singular curves on the right half of the $(q_0, m_0)$ plane lie
between two special accumulation curves, $\gamma(1, 1, 1, 1, ....)$
and $\gamma(1, -1, -1, -1, -1, ...)$ (see Figs. \ref{singular} and
\ref{singularzoom}). So does the fractal of the map on the right
half plane as well (see Fig. \ref{2dmap}). In the whole plane, the
map's fractal lies between the two accumulation curves
$\gamma(1,1,1,\dots)$ and $\gamma(-1,-1,-1,\dots)$.

Similar to the above tracking and identification of singular curves,
we can also track how regions in the $(q, m)$ plane move under the
map $\mathcal{F}$. This is helpful for us to see how the orbit of an
initial point moves in the $(q, m)$ plane. Let us denote the regions
above $\gamma(1,1,1,\dots)$ and below $\gamma(-1,-1,-1,\dots)$ as
${\cal D}_0$, and the region between $\gamma(1, -1, -1, -1, \dots)$
and $\gamma(-1, 1, 1, 1, \dots)$ as ${\cal D}_1$. Then we find that
\begin{equation}
\mathcal{F}^{-1}({\cal D}_0)={\cal D}_0 \cup {\cal D}_1.
\end{equation}
Other regions such as $\mathcal{F}^{-1}({\cal D}_1)$ can be obtained
with the help of singular curves whose pre-images under the inverse
map $\mathcal{F}^{-1}$ have been detailed above (notice that the
singular curve $\gamma_0$ lies in the middle of the region ${\cal
D}_1$). Details will not be pursued in this paper.

\subsection{Asymptotic behaviors of singular curves}

Next we determine the asymptotic behaviors of singular curves as
$q_0 \to 0$ and $\pm \infty$. These asymptotic behaviors will be
needed for our derivation of scaling laws of the exit-velocity
fractals in the ODEs and PDEs as $\varepsilon\to 0$ under the
unequal-amplitude initial conditions (\ref{ic2}), see Sec.
\ref{sec_ic2}. But they will not be needed for scaling laws of
fractals with equal-amplitude initial conditions (\ref{ic1}), see
Sec. \ref{sec_ic1}. Since $\gamma(\mathbf{a})$ and
$\gamma(-\mathbf{a})$ are anti-symmetric to each other, we only
consider curves on the right half plane with
$\mathbf{a}=(1,a_1,...,a_n)$ below.

The asymptotic behaviors of singular curves $\gamma(1,a_1,...,a_n)$
as $q_0\rightarrow0^{+}$ and $+\infty$ can be obtained from
$\gamma(1)$ in Eq. (\ref{gamma1}) and the recursion relations
(\ref{iFiterate1})-(\ref{iFiterate2}), and we get the following
results:
\begin{enumerate}
\item For any singular curve $\gamma(1,\mathbf{a})$, if
$(q_0,m_0)\in\gamma(1, \mathbf{a})$, then
\begin{equation}\label{asym1}
m_0=q_0^{-2}+C_\mathbf{a}+..., \quad q_0\rightarrow0^+,
\end{equation}
where $(2C_\mathbf{a},C_\mathbf{a})$ is the intersection point
between $\gamma(\mathbf{a})$ and $\Omega$.

\item If $(q_0,m_0)\in\gamma(1,a_1,...,a_n)$ with
$a_1=...=a_n=1$ (i.e. on a primary singular curve), then
\begin{eqnarray} \label{asym2a}
m_0=-\frac{1}{2(n+1)}q_0+D_nq_0^{-2}+..., \quad
q_0\rightarrow+\infty,
\end{eqnarray}
where
\begin{equation}
D_n=1+\frac{n+1}{n}D_{n-1}, \quad D_0=1.
\end{equation}

\item If $(q_0,m_0)\in\gamma(1,a_1,...,a_n,-1,\mathbf{\hat{a}})$
where $a_1=...=a_n=1$ and $\mathbf{\hat{a}}$ is an arbitrary finite
binary sequence, i.e. when $(q_0,m_0)$ lies in the secondary
structure of a primary curve $\gamma(1,a_1,...,a_n)$, then
\begin{eqnarray}\label{asym2}
m_0=&&-\frac{1}{2(n+1)}q_0-\frac{1}{\sqrt{2(n+1)}}q_0^{-1/2}\nonumber\\&+&\frac{(n+1)C_\mathbf{\hat{a}}}{\sqrt{2}}q_0^{-3/2}+...,
\quad q_0\rightarrow +\infty.
\end{eqnarray}
where $(2C_\mathbf{\hat{a}},C_\mathbf{\hat{a}})$ is the intersection
point between $\gamma(\mathbf{\hat{a}})$ and $\Omega$.
\end{enumerate}

Notice from (\ref{asym2a}) and (\ref{asym2}) that for any binary
subsequence $\mathbf{\hat{a}}$,
$\gamma(1,a_1,..,a_n,-1,\mathbf{\hat{a}})$ tends to
$\gamma(1,a_1,...,a_n)$ as $q_0\to+\infty$, i.e. the distance
between them goes to zero at large $q_0$ values. This explains why
we could (and should) treat singular curves
$\gamma(1,\mathbf{a},-1,\mathbf{\hat{a}})$ with arbitrary binary
subsequences $\mathbf{\hat{a}}$ as the secondary structures
associated with the primary curve $\gamma(1,\mathbf{a})$ earlier in
this section.

To prove the first asymptotic result (\ref{asym1}), let
\begin{equation}\label{LP-1}
(q_0,m_0)=\mathcal{F}^{-1}(q_1,m_1)=(q_1-2m_1, m_1+\frac{
\mbox{sgn}(q_1-2m_1)}{(q_1-2m_1)^2}).
\end{equation}
Then using (\ref{pmap2}), we get
\begin{equation}\label{P-1}
m_0=\frac{q_1}{2}-\frac{q_0}{2}+\frac{ \mbox{sgn}(q_0)}{q_0^2}.
\end{equation}
For $(q_0,m_0)\in\gamma(1, \mathbf{a})$, in view of the recursion
relation (\ref{iFiterate1}), we see that
$(q_1,m_1)\in\gamma(\mathbf{a})\bigcap\Omega^+$. When $q_0 \to 0^+$,
(\ref{LP-1}) shows that $q_1-2m_1\rightarrow0^+$. Thus $(q_1, m_1)$
approaches the intersection point between $\gamma(\mathbf{a})$ and
$\Omega$, i.e. $q_1\rightarrow 2C_\mathbf{a}$ and $m_1\rightarrow
C_\mathbf{a}$. Substituting these asymptotic results into
(\ref{P-1}), (\ref{asym1}) is then obtained.

The second and third asymptotic relations (\ref{asym2a}) and
(\ref{asym2}) can be proved by the induction method and by utilizing
the recursion relation (\ref{iFiterate1}). These proofs are
elementary and will be omitted.

\section{\label{seODE} Scaling properties of fractals in the ODEs and PDEs}

We have known that fractal structures in the ODEs (\ref{eqDyfinal})
and PDEs (\ref{eqGNLS}) are completely determined by the map's
fractal, together with the initial-value connections
(\ref{scaling1}) and variable scalings (\ref{scaling}) between the
map and the ODEs/PDEs. Utilizing the knowledge we have gained on the
map's fractal in the previous section, we can now obtain a deep
understanding on the fractals in the ODEs and the PDEs. This will be
demonstrated in this section. For definiteness, we will use the two
types of initial conditions (\ref{ic1}) and (\ref{ic2}) as examples.
In both cases, as the initial phase difference $\psi_0$ changes, the
corresponding initial condition of the map forms a parameterized
curve in the $(q_0, m_0)$ plane. This curve, denoted as
$\lambda^\varepsilon$, intersects the map's $|m_\infty|$ fractal in
Fig. \ref{2dmap}, and this intersection then completely determines
the exit-velocity fractals of the ODEs and PDEs shown in Figs.
\ref{comparison} and \ref{comparison2}.

\subsection{The case of equal-amplitude initial conditions}
\label{sec_ic1}

\begin{figure}
\begin{center}
\includegraphics[width=130mm,height=100mm]{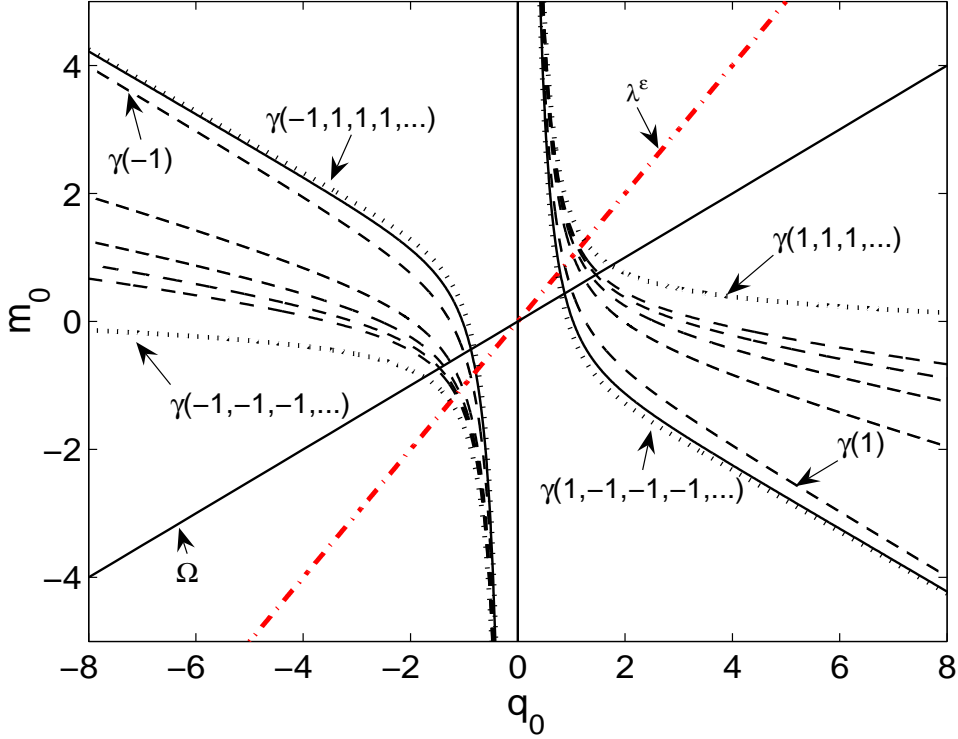}
\caption{\label{iccurve1} Curve $\lambda^\epsilon$ of the map's
initial points (\ref{case1m}) which corresponds to the
equal-amplitude initial conditions (\ref{ic1}). Some singular curves
are also shown. The dashed curves are the same as those in Fig.
\ref{singular}, and the solid curves are $\gamma(1, -1)$ and
$\gamma(-1, 1)$.  }
\end{center}
\end{figure}

In this subsection, we consider the first type of initial conditions
(\ref{ic1}) where the two solitary waves initially have equal
amplitudes. In this case, the corresponding curve of the map's
initial points in the $(q_0, m_0)$ plane can be seen from
(\ref{iv1}) as
\begin{equation}\label{case1m}
\lambda^\varepsilon=\{(q_0,m_0):q_0=m_0=-\frac{\pi^{1/3}}{2\varepsilon^{1/3}}\tan
\psi_0, \; \psi_0\in(-\pi/2,\pi/2) \}.
\end{equation}
The reason for the restriction $\psi_0\in(-\pi/2,\pi/2)$ is that
fractal scatterings can only arise in the $\psi_0$ intervals where
$H<0$ \cite{zhuhabermanyang}. In the present case, $H<0$ corresponds
to the interval $\psi_0\in(-\pi/2,\pi/2)$. This parameterized curve
$\lambda^\varepsilon$ is a straight line with slope one in the
$(q_0,m_0)$ plane, see Fig. \ref{iccurve1}.

The map's $|m_\infty|$ fractal in the initial-condition plane $(q_0,
m_0)$ (see the previous section) is very instrumental for the
understanding of exit-velocity fractals in the PDEs and ODEs. First
of all, from the map's $|m_\infty|$ fractal, together with the
initial-value curve (\ref{case1m}), the formula (\ref{zetaM}) and
various scalings, we can easily construct the map's exit-velocity
fractal in Fig. \ref{comparison}(c). This exit-velocity fractal of
the map can be readily understood. For instance, let us denote the
$\psi_0$ value at the intersection of $\lambda^{\varepsilon}$ with a
singular curve $\gamma(\mathbf{a})$ as
$\psi^{\varepsilon}(\mathbf{a})$. At each
$\psi^{\varepsilon}(\mathbf{a})$, the map's exit-velocity graph has
a singularity peak of infinite height. To illustrate, a few simple
$\psi^{\varepsilon}(\mathbf{a})$ values are marked in Fig.
\ref{comparison}(c). Using this connection, the map's exit-velocity
graph can be completely understood from the map's $|m_\infty|$
fractal. Then the exit-velocity fractals in the ODEs and PDEs can be
similarly understood. To be specific, we find that the primary
window sequence in the exit-velocity fractals of Fig.
\ref{comparison}, which cascades to the left with the first member
being the widest window, are associated with the primary sequence of
singular curves $\{\gamma(1), \gamma(1,1), \dots\}$ (at the
intersection with the set $\lambda^{\varepsilon}$). The dense
secondary structure on the right hand side of each primary window in
the exit-velocity fractals of Fig. \ref{comparison} corresponds to
the secondary structure of each primary-sequence curve in the map's
$|m_\infty|$ fractal. In particular, the value
$\psi^{\varepsilon}(1, -1)$, which is marked in Fig.
\ref{comparison}(c), corresponds to the secondary singular curve
$\gamma(1, -1)$ below the primary curve $\gamma(1)$ in Fig.
\ref{singularzoom}. If we zoom into each secondary structure of a
primary window in the exit-velocity fractals of Fig.
\ref{comparison}, we will see secondary window sequences which
cascade to the right, i.e. the cascading direction of secondary
window sequences is reversed from that of the primary window
sequence. The reason for this is that in the map's $|m_\infty|$
fractal (see Fig. \ref{singular}), the cascading direction of
secondary sequences of singular curves is reversed from that of the
primary sequence as we have explained before. The exit-velocity
fractals of the PDE, the ODE and the map in Fig. \ref{comparison}
can be zoomed further, and all their microscopic structures can be
inferred from the map's singular curves in Fig. \ref{singular}, or
from the map's $|m_\infty|$ fractal in Fig. \ref{2dmap} in general.
One may notice that singularity peaks in the exit-velocity graphs
appear only for the map and the ODEs, but not for the PDEs (see Fig.
\ref{comparison}). Near such singularity peaks, the two solitary
waves collide and coalesce, which makes our reduced ODE model
(\ref{eqDyfinal}) invalid. This explains the difference in those
regions of the exit-velocity graphs between the ODEs and the PDEs.

From the map, we can obtain another important piece of information
on the PDE/ODE's fractal as the initial solitary-wave separation
$\Delta \xi_0$ (i.e. $\zeta_0$) varies. In the present
equal-amplitude initial conditions (\ref{ic1}) and (\ref{icsame}),
if $\Delta \xi_0$ takes other (large) values, we can easily see from
(\ref{case1m}) that the values of $q_0=m_0$ are independent of
$\Delta \xi_0$. This means that the exit-velocity fractal of the map
[see Fig. \ref{comparison}(c)] will remain the same for different
initial solitary-wave separations, which in turn implies the same
for the exit-velocity fractals in the PDE/ODEs. This is a surprising
fact, and it has been confirmed by our direct PDE/ODE simulations.
It is noted that this fact does not hold for the unequal-amplitude
initial conditions (\ref{ic2}) because $(q_0, m_0)$ in such cases
will depend on $\Delta \xi_0$, see (\ref{iv2}).

\begin{figure}
\includegraphics[width=65mm,height=50mm]{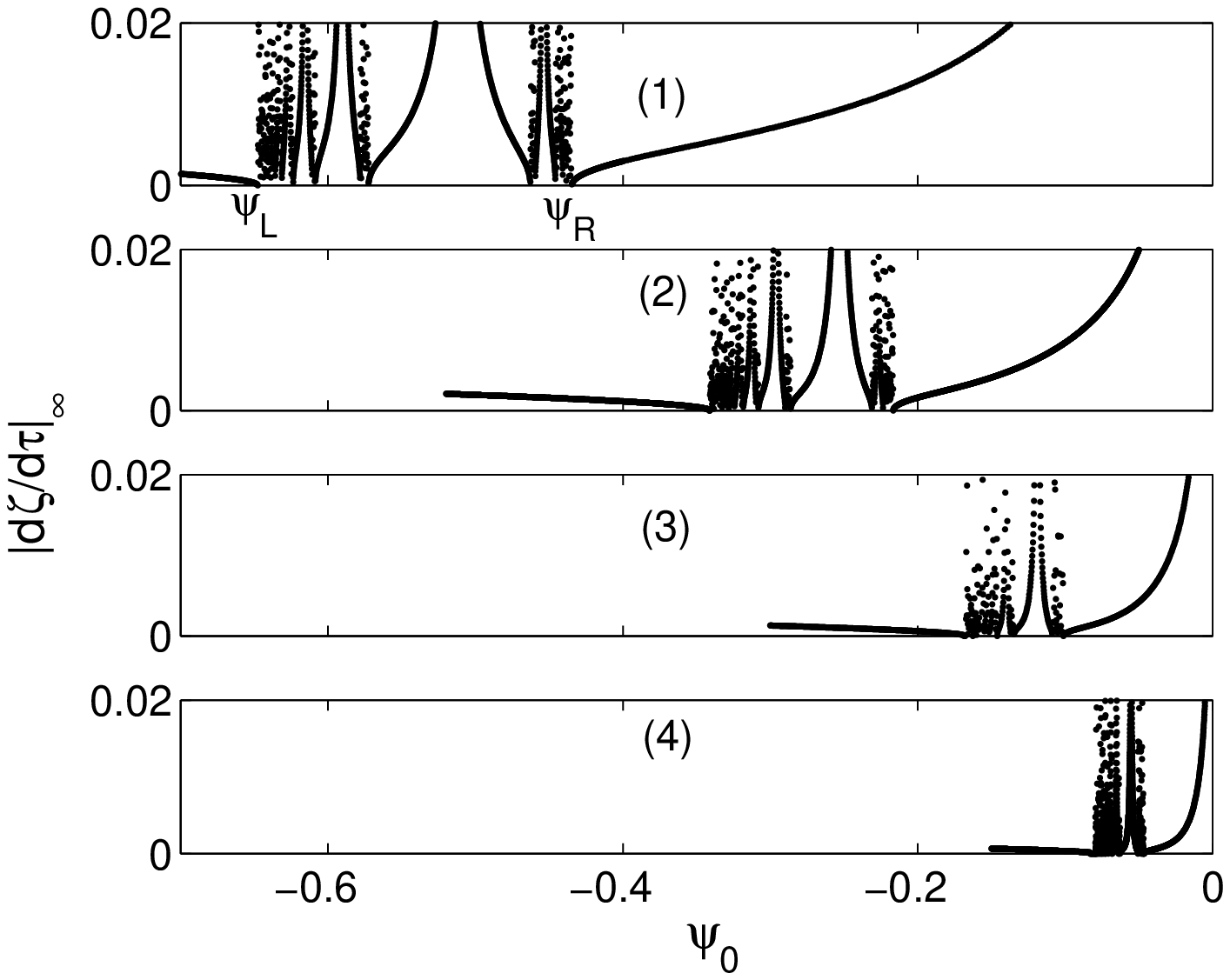}
\includegraphics[width=65mm,height=50mm]{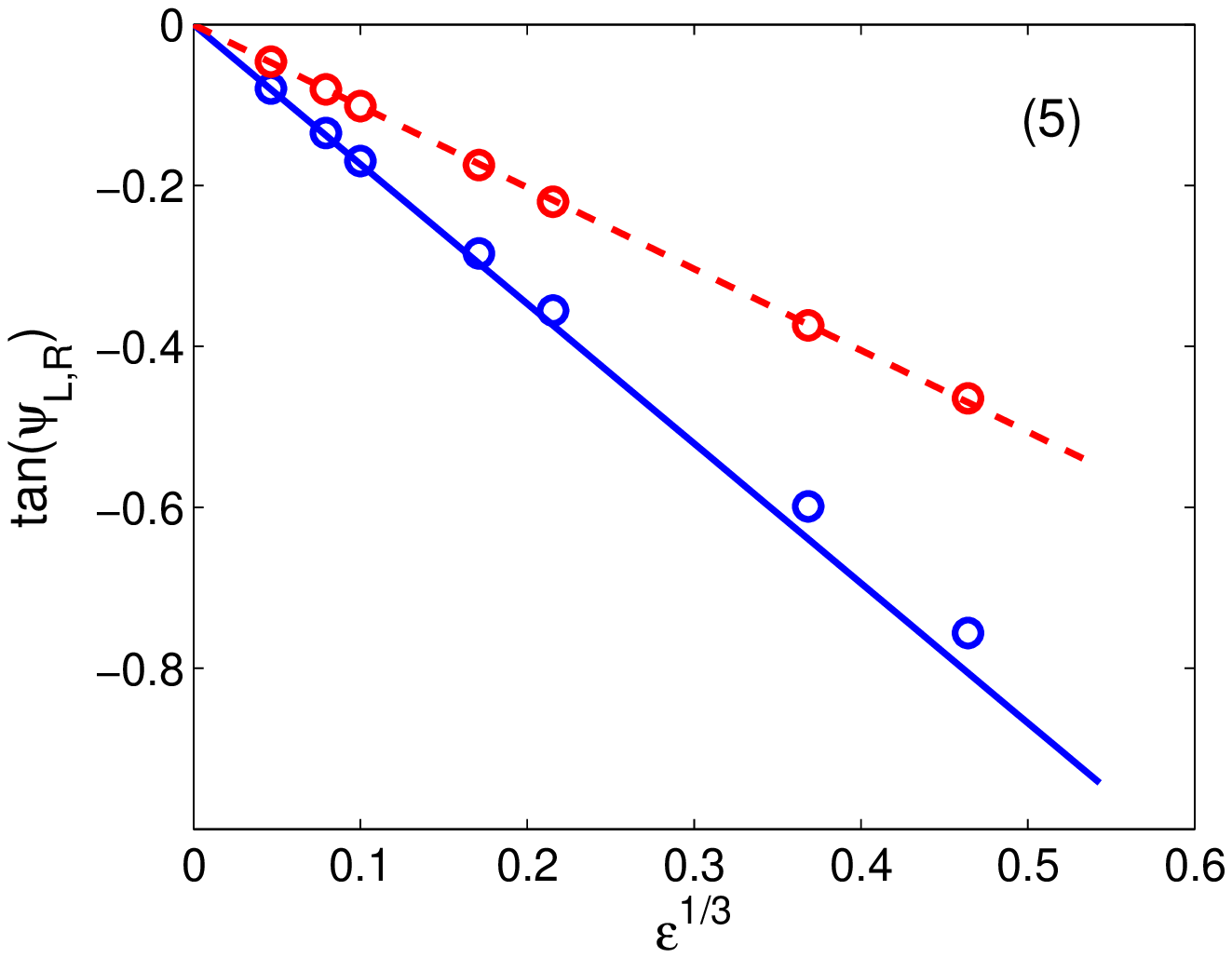}
\caption{\label{shift} Left: change of fractal structures in the
ODEs (\ref{eqDyfinal}) as $\varepsilon\rightarrow 0^+$ under the
equal-amplitude initial condition (\ref{ic1a}). The $\varepsilon$
values are (1) 0.1; (2) 0.01; (3) 0.001 and (4) 0.0001 respectively.
Labels $\psi_L$ and $\psi_R$ represent the left and right boundaries
of the fractal region. Right: $\tan(\psi_L)$ (blue) and
$\tan(\psi_R)$ (red) versus $\varepsilon^{1/3}$. The circles are
data from the ODE simulations, and the straight lines are the
analytical formulae (\ref{phiLR1}) from the map.  }
\end{figure}

In addition to the above qualitative understanding of the
exit-velocity fractals in the PDEs and ODEs, we can further obtain
the scaling properties of these fractals, i.e. we can determine
quantitatively how the fractal structures in the PDEs and ODEs
change as the parameter $\varepsilon$ varies. For this purpose, we
notice that the map's $|m_\infty|$ fractal at the intersection with
$\lambda^{\varepsilon}$ on the right half plane lies between two
accumulation points, $(q_a, q_a)=(0.741, 0.741)$ and $(q_b,
q_b)=(1.271,1.271)$ on $\gamma(1,-1,-1,-1,...)$ and
$\gamma(1,1,1,1,...)$ respectively (see Fig. \ref{iccurve1}). In
view of the initial-condition connection (\ref{case1m}), the
corresponding $\psi_0$ values of these two accumulation points are
\begin{equation} \label{phiLR1}
\psi_L=-\mbox{atan}(2q_b\pi^{-1/3}\varepsilon^{1/3}), \quad
\psi_R=-\mbox{atan}(2q_a\pi^{-1/3}\varepsilon^{1/3}).
\end{equation}
These $\psi_L$ and $\psi_R$ values are the left and right boundaries
of the map's exit-velocity fractal on the negative $\Delta \phi_0$
axis [see Fig. \ref{comparison}(c)], and they are the map's
predictions for the fractal regions in the PDEs/ODEs. These formulae
show that $\psi_{L, R}\rightarrow 0$ as $\varepsilon\rightarrow
0^+$, which means that the whole fractal region shrinks to
$\psi_0=0$ as $\varepsilon\rightarrow0^+$. Notice that when
$\varepsilon=0$, the ODE solution $\zeta(\tau)$ under the
equal-amplitude initial conditions (\ref{ic1a}) develops finite-time
singularity at $\psi_0=0$, where $\mbox{Re}(C_0)=0,
\mbox{Im}(F_0)=0$ (see Sec. \ref{sec2}). Thus when
$\varepsilon\rightarrow0^+$, the fractal region shrinks to the
$\psi_0$ point which develops finite-time singularity in the
integrable ODEs (see also \cite{zhuyang}). Formulae (\ref{phiLR1})
further show that this shrinking is at the rate of
$\varepsilon^{1/3}$. To confirm this analytical prediction, we
directly computed the exit-velocity fractals of the ODEs under the
equal-amplitude initial conditions (\ref{ic1a}) as $\varepsilon$
takes on smaller and smaller values of 0.1, 0.01, 0.001 and 0.0001,
and the results are displayed in Fig. \ref{shift}. We see that as
$\varepsilon\rightarrow0^+$, the ODE's fractal region indeed
approaches $\psi_0=0$ [see Fig. \ref{shift}(1-4)]. In addition, the
fractal region's left and right boundaries $\psi_L$ and $\psi_R$
indeed shrink in proportion to $\varepsilon^{1/3}$. Furthermore, the
constants of proportion match the analytical values in Eq.
(\ref{phiLR1}) as well [see Fig. \ref{shift}(5)]. Similar agreement
has also been found for PDE fractals, see \cite{zhuhabermanyang2}.

The comparison in Fig. \ref{shift}(5) indicates that for the present
equal-amplitude initial conditions, the map's predictions are
asymptotically accurate as $\varepsilon\to 0^+$. This is not
surprising, as the fractal region here lies near $\psi_0=0$ when
$|\varepsilon| \ll 1$. In this case, we can easily see from the
definitions (\ref{Edef}) and (\ref{Mdef}) that $E_0\ll 1$, $M_0\ll
1$ and $M_0/E_0\ll 1$ for the present initial conditions when
$\psi_0$ lies inside the fractal region. Thus the assumptions for
the derivation of the map (\ref{map1})-(\ref{map2}) are satisfied,
and consequently the map's predictions are asymptotically accurate.
For unequal-amplitude initial conditions (\ref{ic1b}), however, the
assumption $M_0/E_0\ll 1$ will not be met in general, thus the map's
predictions will not be asymptotically accurate (even though they
are still qualitatively accurate), see the next section for details.

\subsection{The case of unequal-amplitude initial conditions}
\label{sec_ic2}

\begin{figure}
\begin{center}
\includegraphics[width=130mm,height=100mm]{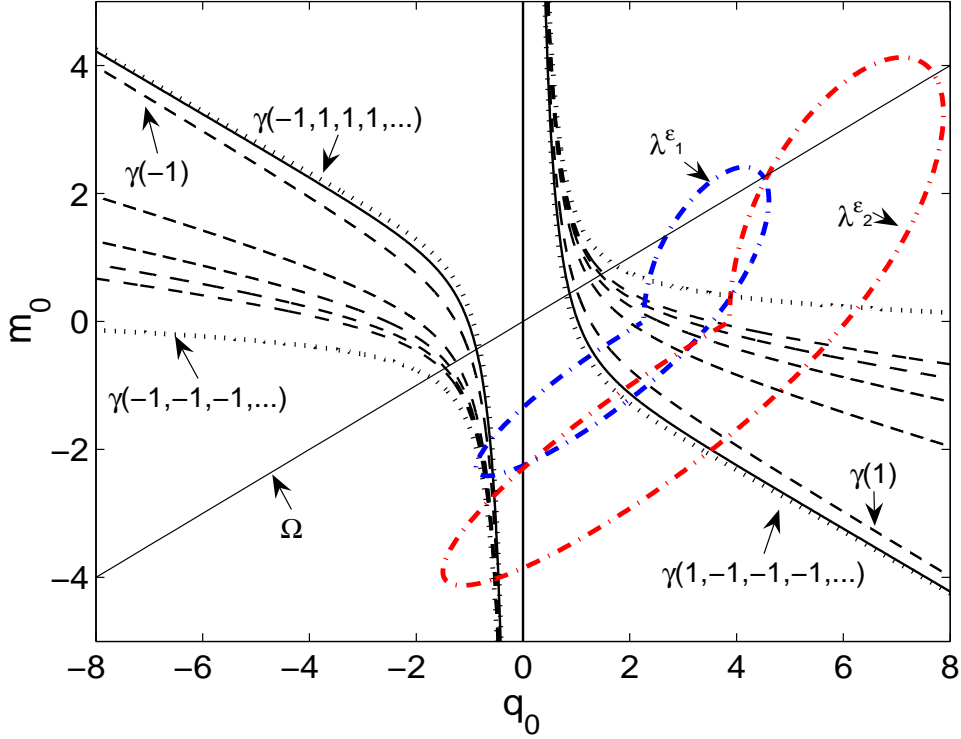}
\caption{\label{iccurve2} Curves of the map's initial points
(\ref{case2m}) corresponding to the unequal-amplitude initial
conditions (\ref{ic1b}) at two $\varepsilon$ values:
$\varepsilon_1=0.01$ (blue) and $\varepsilon_2=0.005$ (red). Some
singular curves are also shown by the same notations as in Fig.
\ref{iccurve1}. }
\end{center}
\end{figure}

Now we consider the second case of unequal-amplitude initial
conditions (\ref{ic2}). In this case, the corresponding curve of the
map's initial values in the $(q_0, m_0)$ plane can be seen from
(\ref{iv2}) as
\begin{equation} \label{case2m}
\lambda^{\varepsilon}=\{(q_0,m_0):
q_0=-T(\psi_0)S(\psi_0)\varepsilon^{-1/3}, \;
m_0=T(\psi_0)\varepsilon^{-1/3}, \; \psi_0\in [0, 2\pi]\},
\end{equation}
where $T(\psi_0)$ is defined in (\ref{Tdef}), $S(\psi_0)$ given by
(\ref{S}), and the other involved parameters specified in
(\ref{ic1b}).
Here $H<0$ in the entire interval of $\psi_0\in [0, 2\pi]$, thus no
restriction on $\psi_0$ is needed \cite{zhuhabermanyang}. These
curves at two $\varepsilon$ values $\varepsilon_1=0.01$ and
$\varepsilon_2=0.005$ are displayed in Fig. \ref{iccurve2}. Each
$\lambda^{\varepsilon}$ is a closed curve, and it intersects a
singular curve $\gamma(\mathbf{a})$ twice. Let us denote the
$\psi_0$ values at the two intersections as
$\psi^{\varepsilon}_1(\mathbf{a})$ and
$\psi^{\varepsilon}_2(\mathbf{a})$, with
$\psi^{\varepsilon}_1(\mathbf{a})<
\psi^{\varepsilon}_2(\mathbf{a})$. These $\psi^{\varepsilon}_{1,
2}(\mathbf{a})$ values are the singularity peaks of infinite height
in the exit-velocity fractal of the map [see Fig.
\ref{comparison2}(c)]. On the singular curve $\gamma_0$, the
intersection points would be denoted as $\psi^{\varepsilon}_1$ and
$\psi^{\varepsilon}_2$. As $\varepsilon \to 0^+$, these
intersections are such that $q_0 \to +\infty$ in the right half
plane and $q_0 \to 0^-$ in the left half plane. This behavior can be
readily understood by examining the intersections of
$\lambda^{\varepsilon}$ with the horizontal and vertical axes. The
intersections with the horizontal axis are such that $m_0=0$. In
view of (\ref{case2m}) as well as the expression (\ref{Tdef}) for
$T(\psi_0)$, we see that these intersections occur when
$T(\psi_0)=0$, i.e. when $\psi_0=0$ and $\pi$, regardless of the
$\varepsilon$ values. Note that under the present initial
conditions,
\begin{equation}
q_A\equiv -\lim_{\psi_0\to 0}T(\psi_0)S(\psi_0)=0.962, \quad
q_B\equiv -\lim_{\psi_0\to \pi}T(\psi_0)S(\psi_0)=0.641,
\end{equation}
thus these two intersections on the horizontal axis are $(q_A
\varepsilon^{-1/3}, 0)$ and $(q_B \varepsilon^{-1/3}, 0)$, which
move to $(+\infty, 0)$ as $\varepsilon\to 0^+$. Similarly, the
intersections of $\lambda^{\varepsilon}$ with the vertical axis
occur when $S(\psi_0)=0$, i.e. when $\psi_A=1.988$ and
$\psi_B=2.871$, regardless of the $\varepsilon$ values. Then the two
intersection points on the vertical axis are $[0,
T(\psi_A)\varepsilon^{-1/3}]$ and $[0,
T(\psi_B)\varepsilon^{-1/3}]$, which approach $(0, -\infty)$ as
$\varepsilon\to 0^+$.

With the help of the map's singular curves as well as the
initial-condition curve $\lambda^{\varepsilon}$ in Fig.
\ref{iccurve2}, we can now understand the map's exit-velocity
fractal in Fig. \ref{comparison2}(c), and hence the PDE/ODEs'
exit-velocity fractals in Fig. \ref{comparison2}(a, b). For these
initial conditions, $\varepsilon=0.001$ (see Sec. 2), and the
initial-value curve $\lambda^{\varepsilon}$ goes outside the box of
Fig. \ref{iccurve2} (thus not displayed). Instead, we will use the
curve $\lambda^{\varepsilon_2}$ in Fig. \ref{iccurve2} (with
$\varepsilon_2=0.005$) as a qualitative guide. The curve
$\lambda^{\varepsilon}$ in the first quadrant [above the
accumulation curve $\gamma(1, 1, \dots)$] corresponds roughly to
$\psi_0\in (\pi, 2\pi)$. This segment of $\lambda^{\varepsilon}$
does not intersect with any singular curves, thus its corresponding
exit-velocity graph would be smooth (in Fig. \ref{comparison2}, this
segment of the graph is not shown). The right intersection point of
$\lambda^{\varepsilon}$ with $\gamma(1, 1, \dots)$ is where
$\psi^{\varepsilon}_1(1, 1, \dots)\approx 0$. This intersection
corresponds to the left edge of the map's exit-velocity fractal in
Fig. \ref{comparison2}(c). From this intersection point down
(leftward), $\lambda^{\varepsilon}$ passes through the primary
sequence of singular curves (in the reverse order), which
corresponds to the primary sequence of singularity peaks starting
from $\Delta\phi_0\approx 0$ rightward (in the reverse order) in
Fig. \ref{comparison2}(c). The singularity peaks
$\psi^{\varepsilon}_1(1)$ and $\psi^{\varepsilon}_1(1, 1)$ in this
primary sequence are marked in Fig. \ref{comparison2}(c), which
correspond to lower intersections of $\lambda^{\varepsilon}$ with
primary singular curves $\gamma(1)$ and $\gamma(1,1)$. At the lower
intersection of $\lambda^{\varepsilon}$ with the vertical axis,
$\psi^{\varepsilon}_1=\psi_A=1.988$, which is labeled in Fig.
\ref{comparison2}(c). From this lower vertical intersection point
leftward, $\lambda^{\varepsilon}$ passes through a thick band of
singular curves, which corresponds to the structures right after the
peak of $\psi^{\varepsilon}_1$ in Fig. \ref{comparison2}(c) (these
structures are not well resolved and only a few vertical points are
visible). After this thick band, $\lambda^{\varepsilon}$ turns
around (upward) and passes through another thick band of singular
curves in the left half plane, which corresponds to the structures
between the label `(c)' and the first major peak to its right in
Fig. \ref{comparison2}(c) (again these structures are not well
resolved). The curve $\lambda^{\varepsilon}$ crosses the vertical
axis again (upper intersection) at
$\psi^{\varepsilon}_2=\psi_B=2.871$, which is labeled in Fig.
\ref{comparison2}(c). From this upper vertical intersection
rightward, $\lambda^{\varepsilon}$ passes through the primary
sequence of singular curves again (in forward order), which
corresponds to the primary sequence of singularity peaks from
$\psi^{\varepsilon}_2$ rightward and ending at
$\psi^{\varepsilon}_2(1, 1, \dots)\approx \pi$ in Fig.
\ref{comparison2}(c). From this correspondence between the
initial-value curve $\lambda^{\varepsilon}$ and the map's
exit-velocity fractal in Fig. \ref{comparison2}(c), a good and clear
understanding of the exit-velocity fractals in the PDEs/ODEs [see
Fig. \ref{comparison2}(a, b)] is then reached.

From the previous section, we know that when $q_0\to +\infty$, the
map's secondary singular curves below each primary singular curve
approach this primary curve. In addition, it is easy to see that all
singular curves in the left half plane approach $\gamma_0$ (the
vertical axis) when $q_0\to 0^-$. Then in view of the
small-$\varepsilon$ asymptotics of curve $\lambda^{\varepsilon}$
described above, we see that as $\varepsilon\to 0^+$, the
intersections of $\lambda^{\varepsilon}$ with the map's singular
curves approach the primary sequence $\{\gamma_0, \gamma(1),
\gamma(1, 1), \gamma(1,1,1), \dots\}$. As a result, when
$\varepsilon\to 0^+$, the exit-velocity fractals in the PDEs/ODEs
will converge to certain discrete $\psi_0$ values whose
corresponding $(q_0, m_0)$ points fall on the above primary
sequence. As was explained in \cite{zhuhabermanyang}, such discrete
$\psi_0$ values are precisely the initial-condition points whose
$\zeta$-orbit in the integrable ODE system (\ref{eqDyfinal})
develops finite-time singularities (this fact will be re-established
again later in this section).

In addition to the above qualitative descriptions of the PDE/ODE's
fractals (see Fig. \ref{comparison2}), we can further determine how
these fractal structures change quantitatively as $\varepsilon \to
0^+$. For instance, we can determine how the singularity peaks in
the ODEs' exit-velocity fractal of Fig. \ref{comparison2}(b) move as
$\varepsilon$ varies [the PDE's fractal does not have singularity
peaks but only counterpart structures, see Fig.
\ref{comparison2}(a)]. This can be done because we know how the
curve $\lambda^{\varepsilon}$ moves with $\varepsilon$ [see Eq.
(\ref{case2m}) and Fig. \ref{iccurve2}]. In addition, its
intersections with singular curves of the map have either $q_0\to
+\infty$ (in the right half plane) or $q_0\to 0^-$ (in the left half
plane), where the asymptotics of singular curves have been derived
in the previous section. Thus when the $(q_0, m_0)$ relation of the
curve $\lambda^{\varepsilon}$ in (\ref{case2m}) is inserted into the
asymptotic equations of singular curves and all variables are
expressed in terms of the control parameter $\psi_0$, quantitative
changes of singularity peaks in the ODEs' exit-velocity fractals
will be obtained. After simple algebra, we get the following
asymptotic results in the limit of $\varepsilon\to 0^+$:
\begin{equation} \label{Pscaling0}
\psi^{\varepsilon}_1=\psi_A, \quad \psi^{\varepsilon}_2=\psi_B,
\end{equation}
\begin{equation} \label{Pscaling}
\psi^{\varepsilon}_k(1,\mathbf{a})=\psi^{0}_k(1,\mathbf{a})+A_k(\mathbf{a})\varepsilon+\dots,
\end{equation}
\begin{equation} \label{Sscaling}
\psi^{\varepsilon}_k(1,\mathbf{a},-1,\mathbf{\hat{a}})=\psi^{0}_k(1,\mathbf{a})+G_k(\mathbf{a})
\varepsilon^{1/2}+H_k(\mathbf{a},
\mathbf{\hat{a}})\varepsilon^{5/6}+\dots,
\end{equation}
\begin{equation} \label{PLscaling}
\psi^{\varepsilon}_k(-1,\mathbf{\hat{a}})=\psi^{0}_k +P_k
\varepsilon^{1/2}+Q_k(\mathbf{\hat{a}})\varepsilon^{5/6}+\dots.
\end{equation}
Here $\mathbf{a}=(a_1,...,a_n)$, $a_1=...=a_n=1$, $\mathbf{\hat{a}}$
is an arbitrary finite binary sequence, $A_k, G_k, H_k, Q_k$ are
constants whose values depend on the binary sequences behind them,
$P_k$ are $\mathbf{\hat{a}}$-independent constants, and $k=1, 2$. If
$n=0$, then $\mathbf{a}$ is empty (which is allowed). Eq.
(\ref{Pscaling0}) indicates that the singularity points
$\psi^{\varepsilon}_k$ in the ODEs' fractal [see Fig.
\ref{comparison2}(b)] are $\varepsilon$-independent. Actually these
$\psi^{\varepsilon}_k$ points in the ODEs/PDEs do depend weakly on
$\varepsilon$, but this weak $\varepsilon$-dependence can not be
captured by our map since it is beyond the asymptotic validity of
the map. Eq. (\ref{Pscaling}) describes how a primary singularity
point approaches its integrable counterpart as $\varepsilon\to 0^+$,
and this convergence is at the uniform rate of $O(\varepsilon)$.
Relations (\ref{Sscaling}) and (\ref{PLscaling}) describe how the
secondary structure of a primary singularity point approaches the
integrable counterpart of this singularity point, and this
convergence is at the uniform rate of $O(\varepsilon^{1/2})$. The
length of the secondary structure, on the other hand, shrinks at the
rate of $O(\varepsilon^{5/6})$. By substituting the $(q_0, m_0)$
relation of (\ref{case2m}) into the asymptotic equations
(\ref{asym2a}) of primary curves and taking the limit of
$\varepsilon\to 0$, we see that
\begin{equation} \label{S2n}
S(\psi^{0}_k(1,\mathbf{a}))=2(n+1), \quad S(\psi^{0}_k)=0, \quad
k=1, 2.
\end{equation}
In addition, at these $\psi^{0}_k$ and $\psi^{0}_k(1,\mathbf{a})$
values, it is easy to check that $\mbox{Re}(C_0)\ne 0$, thus
solutions of the integrable ODEs develop finite-time singularities
in $\zeta(\tau)$ (see Sec. 2 and \cite{zhuhabermanyang}). Thus
relations (\ref{Sscaling}) and (\ref{PLscaling}), together with Eq.
(\ref{phiLR1}) and the discussions below it, quantitatively prove
that when $\varepsilon \to 0^+$, the fractal regions in the
non-integrable ODEs shrink to the $\psi_0$ points which develop
finite-time singularities in the integrable ODEs, as was originally
observed numerically in \cite{zhuyang}.

\begin{figure}
\includegraphics[width=65mm,height=60mm]{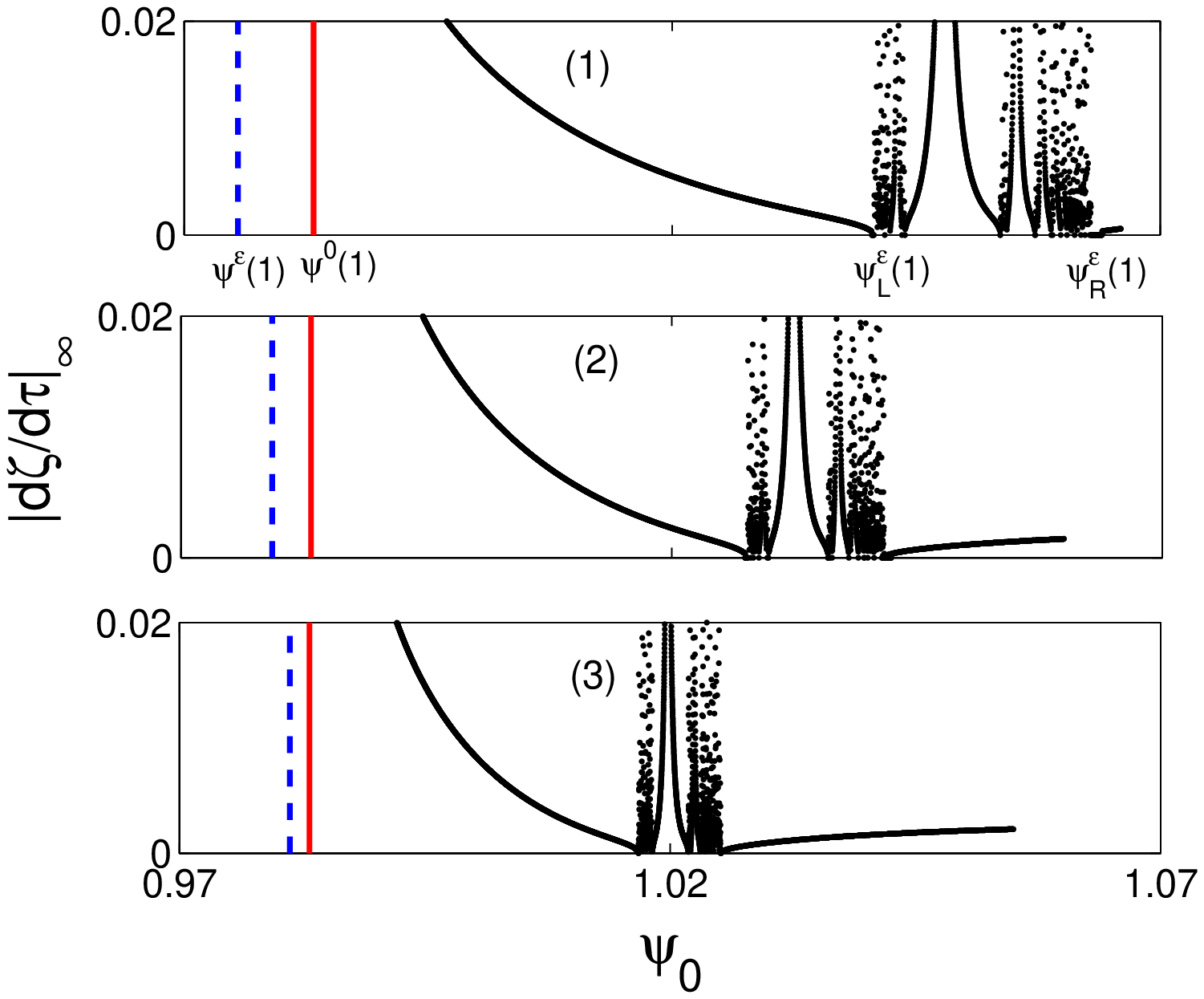}
\includegraphics[width=65mm,height=60mm]{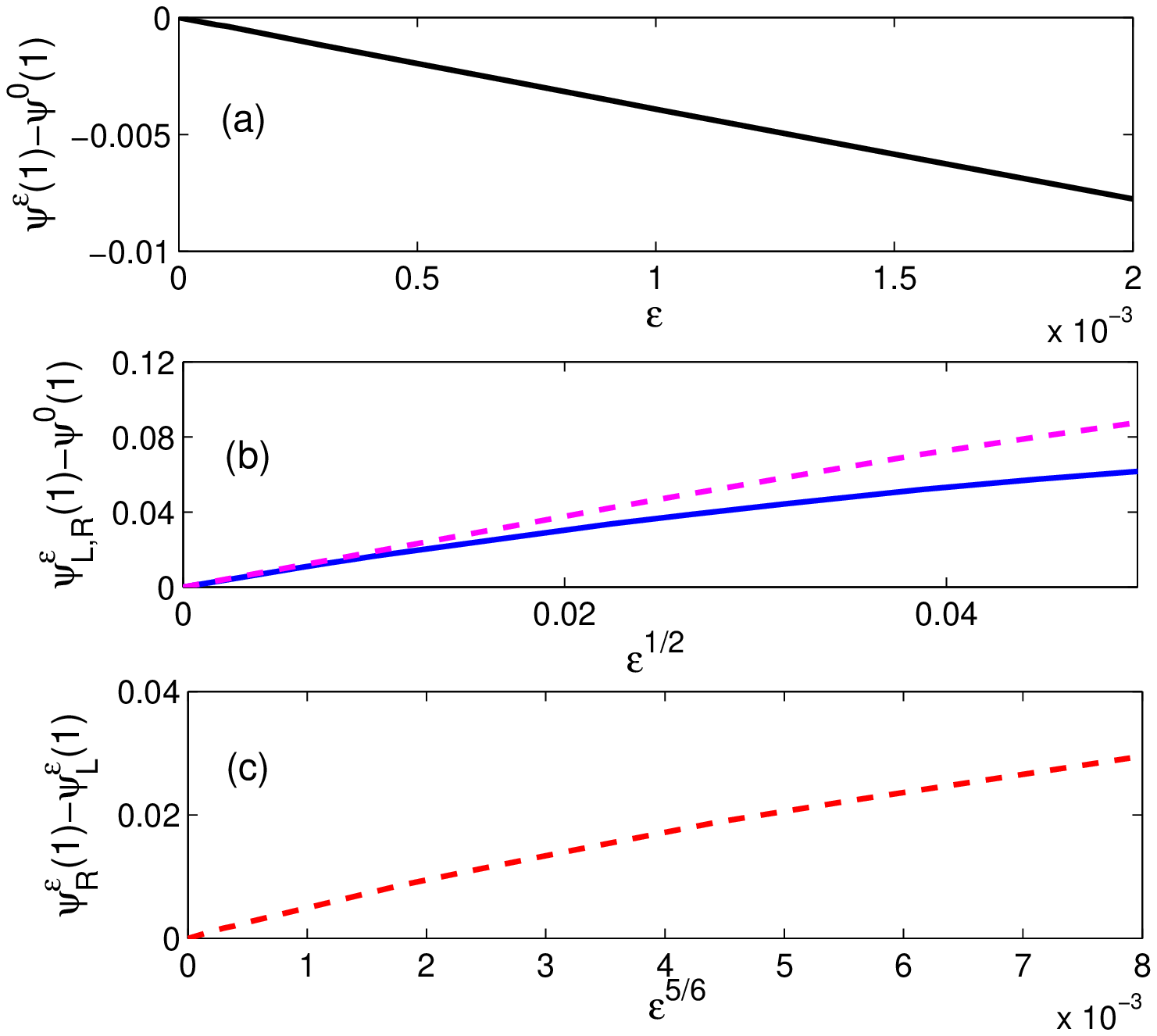}
\caption{\label{un2zero} Left: segments of the ODE's fractal
structures near the lower intersection between
$\lambda^{\varepsilon}$ and the primary singular curve $\gamma(1)$
under unequal-amplitude initial conditions (\ref{ic1b}) [this
segment for $\varepsilon=0.001$ was marked in the whole fractal
structure of Fig. \ref{comparison2}(b)]. The $\varepsilon$ values
are (1) 0.002; (2) 0.001 and (3) 0.0005. Right: (a)
$\psi^{\varepsilon}(1)-\psi^0(1)$ versus $\varepsilon$; (b)
$\psi^{\varepsilon}_L(1)-\psi^0(1)$ (blue) and
$\psi^{\varepsilon}_R(1)-\psi^0(1)$ (red) versus
$\varepsilon^{1/2}$; (c)
$\psi^{\varepsilon}_R(1)-\psi^{\varepsilon}_L(1)$ versus
$\varepsilon^{5/6}$. }
\end{figure}

Now we compare the above analytical $\varepsilon$-scaling laws for
the exit-velocity fractals in the PDEs/ODEs with direct numerical
simulation results. Comparisons with only ODE simulations will be
performed, as PDE simulations at very small $\varepsilon$ values are
very expensive and time consuming. For the ease of comparison, we
take a small segment of the exit-velocity fractal in the ODEs as
marked in Fig. \ref{comparison2}(b), which corresponds to the
segment on the initial-value curve $\lambda^{\varepsilon}$
containing the lower intersection with $\gamma(1)$ and its secondary
structure. Then we monitor how this segment of the fractal moves as
$\varepsilon$ varies by directly simulating the ODEs
(\ref{eqDyfinal}). At three $\varepsilon$ values of 0.002, 0.001 and
0.0005, these segments of the fractal structures in the ODEs are
displayed in Fig. \ref{un2zero} (1-3) respectively. Here the
vertical solid lines mark the singularity point $\psi^0(1)$ in the
integrable ODE system, the vertical dashed lines mark the
singularity point $\psi^\varepsilon(1)$ in the non-integrable ODEs
(with non-zero $\varepsilon$), and labels $\psi_{L,
R}^\varepsilon(1)$ mark the left and right ends of the secondary
structure. It is seen from these figures that as $\varepsilon\to
0^+$, both $\psi^\varepsilon(1)$ and the secondary structure
approach the singularity point $\psi^0(1)$ of the integrable ODEs
(as predicted). To determine the convergence rates, the graphs of
$\psi^{\varepsilon}(1)-\psi^0(1)$, $\psi_{L,
R}^{\varepsilon}(1)-\psi^0(1)$ and
$\psi^{\varepsilon}_R(1)-\psi^{\varepsilon}_L(1)$ at various values
of $\varepsilon$ are plotted in Fig. \ref{un2zero}(a, b, c)
respectively. It is seen that as $\varepsilon\to 0^+$,
$\psi^{\varepsilon}(1)-\psi^0(1)$ approaches a linear function in
$\varepsilon$, confirming the $O(\varepsilon)$-convergence of the
primary singularity point $\psi^{\varepsilon}(1)$ toward the
integrable singularity point $\psi^0(1)$ [see formula
(\ref{Pscaling})]. Quantities $\psi_{L,
R}^{\varepsilon}(1)-\psi^0(1)$ approach a linear function in
$\varepsilon^{1/2}$, confirming the
$O(\varepsilon^{1/2})$-convergence of the secondary structure toward
the integrable singularity point $\psi^0(1)$  [see formula
(\ref{Sscaling})]. In particular, it is seen that as $\varepsilon\to
0^+$, $\psi_L^{\varepsilon}(1)-\psi^0(1)$ and
$\psi^{\varepsilon}_R(1)-\psi^0(1)$ approach linear functions in
$\varepsilon^{1/2}$ with the same slope, confirming the formula
(\ref{Sscaling}) that the coefficient $G_k$ of $\varepsilon^{1/2}$
is independent of the singular curves inside the secondary
structure. The quantity
$\psi^{\varepsilon}_R(1)-\psi^{\varepsilon}_L(1)$ approaches a
linear function in $\varepsilon^{5/6}$, confirming the
$O(\varepsilon^{5/6})$ shrinking rate of the length of the secondary
structure [see (\ref{Sscaling})]. Thus our analytical predictions
(\ref{Pscaling})-(\ref{PLscaling}) on the order of convergence of
ODE fractals as $\varepsilon\to 0^+$ are confirmed.

Quantitatively, the coefficients in front of the order of
convergence in the analytical formulae
(\ref{Pscaling})-(\ref{PLscaling}) do not match numerical values
though. For instance, in Fig. \ref{un2zero}(a), the slope of
$\psi^{\varepsilon}(1)-\psi^0(1)$ with respect to $\varepsilon$ is
found numerically to be $-3.89$, which differs from the analytical
value of $A(1)=-5.12$ in formula (\ref{Pscaling}). This means that
in the present case of unequal-amplitude initial conditions
(\ref{ic1b}), our analytical formulae
(\ref{Pscaling})-(\ref{PLscaling}) are not asymptotically accurate
when $\varepsilon \ll 1$, which contrasts the equal-amplitude
initial-condition case in Sec. \ref{sec_ic1} (where our analytical
predictions were asymptotically accurate). The reason for this is
that our map (\ref{map1})-(\ref{map2}), or
(\ref{pmap1})-(\ref{pmap2}), was derived asymptotically under the
condition of $M_n/E_n\ll 1$ \cite{zhuhabermanyang2,zhuhabermanyang}.
This condition was satisfied for the equal-amplitude initial
conditions (\ref{ic1}) (see Sec. \ref{sec_ic1}), but is not
satisfied for the unequal-amplitude initial conditions (\ref{ic1b}).
Indeed, one can easily check that for $\psi_0$ values in the fractal
regions of unequal-amplitude initial conditions (see Figs.
\ref{comparison2} and \ref{un2zero}), $M_0/E_0$ does not tend to
zero as $\varepsilon\to 0$, thus the map's predictions become
asymptotically inaccurate as we have just observed. Qualitatively,
the map's predictions are still correct as Fig. \ref{un2zero} has
demonstrated.

\section{Dynamics of ODE and PDE solutions in the fractal structures}
From the previous two sections, we have reached a clear and deep
understanding on the fractal graphs of exit velocities in the ODEs
and PDEs (see Figs. \ref{comparison} and \ref{comparison2}). In this
section, we explain the solution dynamics of the ODEs and PDEs on
these fractals. Notice that these fractals consist of hills of
various widths. The `center' of each hill corresponds to a singular
curve in the map's fractal, thus each hill can be identified by the
binary sequence of that singular curve. We will show that once the
binary sequence of a hill is given, then the solution dynamics of
the ODEs and PDEs on that hill can be ascertained. Throughout this
section, $\mbox{sgn}(\varepsilon)=1$, where fractal scatterings
occur.

\begin{figure}
\begin{center}
\includegraphics[width=100mm,height=50mm]{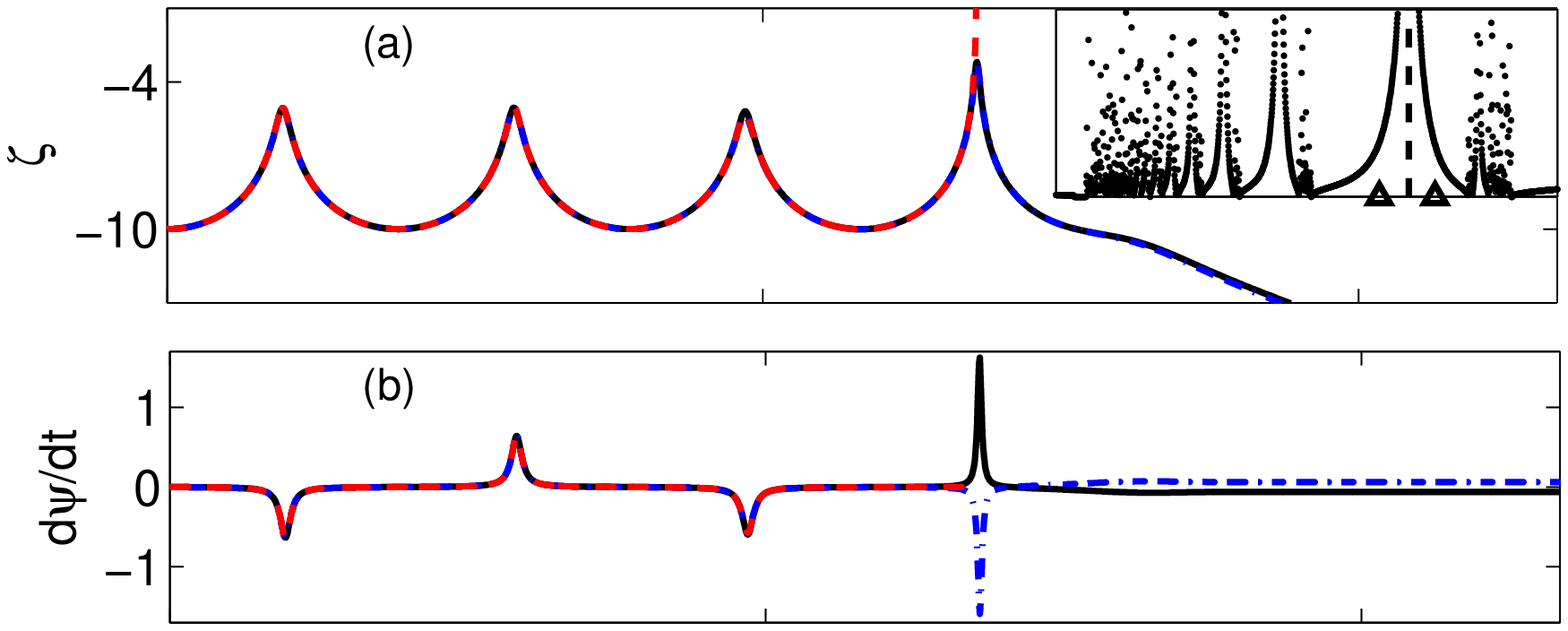}
\includegraphics[width=100mm,height=60mm]{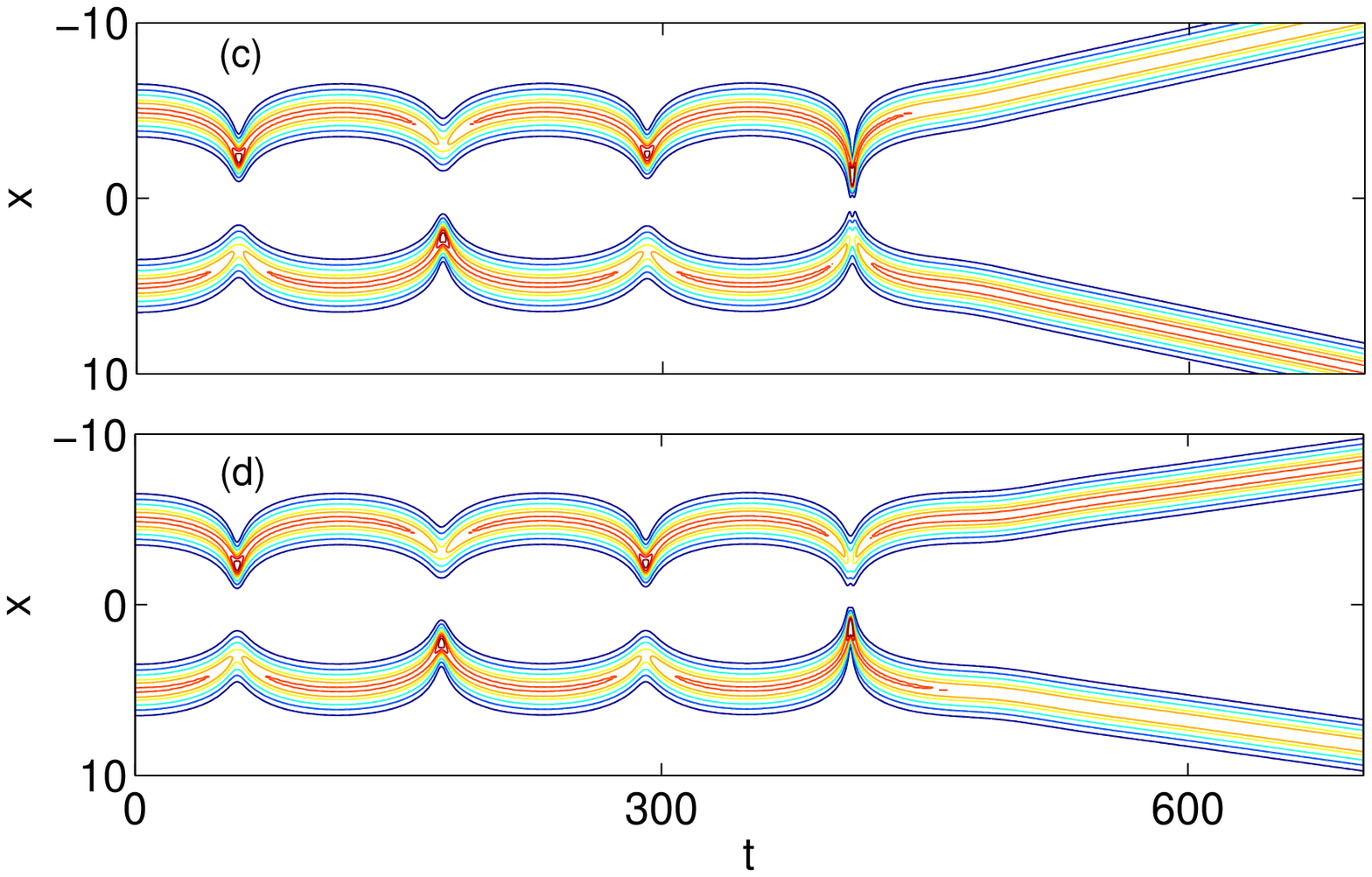}
\end{center}
\caption{\textit{Dynamics of the ODE and PDE solutions on the hill
of binary sequence $\mathbf{a}=(1, -1, 1)$ in the exit-velocity
fractals for equal-amplitude initial conditions (\ref{ic1}). The
inset in (a) is an amplification of the segment marked in the ODE
fractal of Fig. \ref{comparison}(b), where the widest hill is the
one of binary sequence $\mathbf{a}=(1, -1, 1)$. Three locations
$\psi_a<\psi_b<\psi_c$ on this hill are marked by two triangles and
a vertical dashed line, with the dashed line at $\psi_b$ being the
singularity peak point. The ODE solutions at $\psi_a$, $\psi_b$ and
$\psi_c$ are plotted as dash-dotted lines (blue), dashed lines (red)
and solid lines (black) in (a, b) respectively. The PDE solutions at
points corresponding to $\psi_a$ and $\psi_c$ in the PDE fractal of
Fig. \ref{comparison}(a) are shown in (c, d) respectively (contour
plots).  }} \label{dynamics}
\end{figure}

We first describe the ODE solution dynamics in the exit-velocity
fractal. To demonstrate, we pick the hill of binary sequence
$\mathbf{a}=(1, -1, 1)$ in the ODE's fractal for equal-amplitude
initial conditions in Fig. \ref{comparison}(b). The segment which
contains this hill is marked in that fractal. The amplification of
this segment is shown in the inset of Fig. \ref{dynamics}(a), where
the widest hill is the one of this binary sequence $\mathbf{a}$. To
examine the dynamics of ODE solutions on this hill, we pick three
points on the hill with $\psi_a<\psi_b<\psi_c$, where the middle
point $\psi_b$ is the singularity peak point, and the other two
points are on the two sides of this singularity point. These three
points are marked in the inset of Fig. \ref{dynamics}(a). At these
three points, the ODE solutions $\zeta(t)$ and $\dot{\psi}(t)$ are
displayed in Fig. \ref{dynamics}(a, b) respectively. Here the time
$t$ has been rescaled back to the physical time for easy comparison
with the PDE dynamics below. We see that at the singularity peak
point $\psi_b$, the $\zeta(t)$ solution oscillates three times, then
develops finite-time singularity at $t_c \approx 408$ and terminates
there. Each local minimum of $\zeta$ is a saddle approach
\cite{zhuhabermanyang} where the separation between the two waves is
the largest. Each local maximum of $\zeta$ is a `bounce' point where
the two waves are locally the closest and interact more strongly.
The $\dot{\psi}(t)$ solution is mostly flat, except that it exhibits
spikes at the three bounce points whose sign sequence is $(-1, 1,
-1)$, which is opposite of the binary sequence $\mathbf{a}$. This
$\dot{\psi}(t)$ solution also terminates at the finite-time
singularity time $t_c$. At the other two points $\psi_a$ and
$\psi_c$ on the two sides of the singularity peak $\psi_b$, the ODE
solutions do not develop finite-time singularities. From Fig.
\ref{dynamics}(a, b), we see that these solutions are almost
indistinguishable from the singular solution of $\psi_b$ up to the
singularity time $t_c$. At $t\approx t_c$, the $\zeta(t)$ solutions
in both cases have a global maximum, where the two waves are the
closest and interact most strongly. This time was called the
\emph{collision time} in \cite{zhuyang}. Beyond this time, the
$\zeta(t)$ solutions of $\psi_a$ and $\psi_c$ both go to $-\infty$
with finite speed. The $\dot{\psi}$ solutions of $\psi_a$ and
$\psi_c$ approach constants as $t\to \infty$, but these asymptotic
constants have opposite signs:  the asymptotic constant for $\psi_a$
at the left side of the singularity peak is positive, while that for
$\psi_c$ at the right side of the singularity peak is negative. At
other points on this hill, the ODE solution dynamics is
qualitatively similar to the ones above. From these examples, we can
draw general conclusions for the ODE dynamics on a hill of an
arbitrary binary sequence $\mathbf{a}=(a_0, a_1, a_2, \dots, a_n)$
in the exit-velocity fractal. At the singularity peak of the hill,
the $\zeta(t)$ solution oscillates $n+1$ times (i.e. the two waves
bounce with each other $n+1$ times), and then approaches infinity at
a finite time $t_c$ and terminates. Here $n+1$ is the length of the
binary sequence $\mathbf{a}$. The $-\dot{\psi}$ solution exhibits
spikes at the $n+1$ bounce points before the singularity time $t_c$,
whose sign sequence is $\mathbf{a}$. At other points of the hill,
the ODE solutions are almost indistinguishable from this singular
solution up to the singularity time $t_c$. The time $t_c$ is
approximately the collision time of all these ODE solutions. Beyond
this collision time, all $\zeta(t)$ solutions go to $-\infty$, while
all the $\dot{\psi}(t)$ solutions approach constants which have the
same sign on the same side of the hill but opposite sign between the
two sides.

From the above description of ODE dynamics on hills of exit-velocity
fractals, we see that the `physical' meaning of the binary sequence
$\mathbf{a}$ of a hill in the ODE solutions is that $\mathbf{a}$
gives the sign sequence of the $-\dot{\psi}(t)$ solution at the
bounce points before the collision time. In addition, the difference
in solution dynamics between the left and right sides of the hill is
that the asymptotic constants of their $-\dot{\psi}(t)$ solutions
have opposite signs. These two facts can be readily explained.
First, from the definition (\ref{gammaa}), we know that the binary
sequence $\mathbf{a}$ is the signs of $(q_0, q_1, \dots, q_n)$, with
$q_{n+1}=0$. This orbit of the map corresponds to the ODE solution
at the singularity peak of binary sequence $\mathbf{a}$ in the
exit-velocity fractal. From Eqs. (\ref{map1}) and (\ref{scaling}),
we see that $q_k$ has the sign of $-\Delta M_k$, where $\Delta
M_k=M_{k+1}-M_k$, and $M_k$ is the momentum value at the $k$-th
saddle approach. Hence $\mathbf{a}$ is also the signs of $(-\Delta
M_0, -\Delta M_1, \dots, -\Delta M_n)$. Furthermore, from Ref.
\cite{zhuhabermanyang} [see Eqs. (3.4) and (5.6) in particular],
sgn($-\Delta M_k$) is the sign of $-\dot{\psi}$ at the $k$-th
$\zeta$-maximum (bounce point). Thus the binary sequence
$\mathbf{a}$ is equal to the sign sequence of $-\dot{\psi}$ at the
bounce points (before the singularity time or the collision time).
Regarding the signs of $-\dot{\psi}(\infty)$ on the two sides of the
hill, we notice from the map (\ref{pmap1})-(\ref{pmap2}) that in the
map's $|m_\infty|$ fractal in Figs. \ref{2dmap} and \ref{singular},
when $(q_0, m_0)$ is on the left (right) side of the vertical axis
$\gamma_0$ and in its vicinity, $q_\infty$ is positive (negative).
Using the recursive relations (\ref{iFiterate1})-(\ref{iFiterate2})
between singular curves and the orientation-preserving property of
the inverse map $\mathcal{F}^{-1}$, we see that for $(q_0, m_0)$
lying on the left (right) side of every singular curve (and in its
vicinity), $q_\infty$ is positive (negative). Notice that like
bounce points above, the sign of $q_\infty$ is the same as that of
$-\dot{\psi}(\infty)$. In addition, the two sides of a hill in the
ODE's exit-velocity fractal correspond to the two sides of the
singular curve in the map's $|m_\infty|$ fractal. Thus the values of
$-\dot{\psi}(\infty)$ on the two sides of a hill in the
exit-velocity fractal have opposite signs. The specific signs of
$-\dot{\psi}(\infty)$ on the two sides of the hill depend on how
these two sides of the hill correspond to the two sides of the
singular curve. If the left side of the hill corresponds to the left
side of the singular curve, which is the case for hills starting
from $\psi_2^\varepsilon$ rightward in the unequal-initial-amplitude
fractal of Fig. \ref{comparison2}(b), then $-\dot{\psi}(\infty)$
would be positive (negative) on the left (right) side of the hill.
But if the left side of the hill corresponds to the right side of
the singular curve, which is the case for all hills in the
equal-initial-amplitude fractal of Fig. \ref{comparison}(b) and the
hills starting from $\psi_1^\varepsilon$ leftward in the
unequal-initial-amplitude fractal of Fig. \ref{comparison2}(b), then
$-\dot{\psi}(\infty)$ would be negative (positive) on the left
(right) side of the hill.

The topic of higher interest to us is the interaction dynamics in
the PDEs (\ref{eqGNLS}) rather than in the ODEs (\ref{eqDyfinal}).
So next we describe the PDE solution dynamics in the exit-velocity
fractal. This PDE dynamics can be predicted from the ODE dynamics
above. To illustrate, we take two $\Delta \phi_0$ values on the hill
of binary sequence $\mathbf{a}=(1, -1, 1)$ in the PDE's
equal-initial-amplitude fractal of Fig. \ref{comparison}(a), which
correspond to the two $\psi_0$ values of the ODEs on the two sides
of the singularity peak as marked in the inset of Fig.
\ref{dynamics}(a). For these two $\Delta \phi_0$ values, the PDE
solutions are displayed in Fig. \ref{dynamics}(c, d). We see that
before the collision time $t_c\approx 408$, the two PDE solutions
are almost identical with each other. In this time period, the two
waves bounce three times. At the three bounce points, the left wave
has higher, lower and higher amplitudes sequentially (when compared
to the right wave). At the collision time, the two waves are the
closest and interact most strongly. Afterwards, they separate from
each other and escape to infinity. The main difference between the
two PDE solutions is that, at the left $\Delta \phi_0$ value, the
exiting right wave has higher amplitude, while at the right $\Delta
\phi_0$ value, it is just the opposite. These behaviors are common
for most of the points on this hill in the PDE's fractal. The only
exception is a small region in the middle of this hill which
corresponds to the singularity peak point and its immediate vicinity
in the ODE's fractal [see inset of Fig. \ref{dynamics}(a)]. In that
region, the $\zeta$ solutions in the ODEs are either infinite or
very large at the singularity time or collision time, which implies
that the two waves collide and coalesce. When this happens, the
reduced ODE model (\ref{eqDyfinal}) and its predictions become
invalid. Indeed in the PDE's fractal, the central part of every hill
dips down, which contrasts with the ODE's fractal where the central
part of each hill rises up to a peak of infinite height.

The PDE solutions in Fig. \ref{dynamics}(c, d) directly correspond
to the ODE solutions at the two sides of the singularity peak in
Fig. \ref{dynamics}(a, b). In particular, the sign sequence
$\mathbf{a}=(1, -1, 1)$ of the ODE's $-\dot{\psi}(t)$ solution at
the three bounce points directly implies the ``higher, lower,
higher" amplitudes of the left wave at the three bounce points in
the PDE solution, and the negative (positive) $-\dot{\psi}(\infty)$
values at the left (right) side of the singularity peak directly
implies the lower (higher) amplitude of the exiting left wave in the
PDE solution. These connections can be readily explained. Let us
recall the relation $\Delta\phi_t=\Delta \beta$ as well as the fact
that the propagation constant $\beta$ is directly related to the
wave amplitude \cite{zhuyang}. Then the sign of $-\dot{\psi}$, which
is equal to the sign of $-\Delta \beta=\beta_1-\beta_2$, tells which
of the two waves has higher amplitude. For the cubic-quintic
nonlinearity (\ref{cqn}), the amplitude is an increasing function of
$\beta$, thus positive $-\dot{\psi}$ means that the left wave has
higher amplitude, which explains the above connections.

Based on the above PDE examples and general ODE dynamics, we can
draw general conclusions for the PDE dynamics on a hill of an
arbitrary binary sequence $\mathbf{a}=(a_0, a_1, a_2, \dots, a_n)$
in the exit-velocity fractal. For all points on the hill, the PDE
solutions are almost identical to each other up to the collision
times (whose values are almost the same for the entire hill). Before
the collision time, the two waves bounce $n+1$ times. At each bounce
point, the left wave has higher (lower) amplitude if the
corresponding digit in the binary sequence $\mathbf{a}$ is $1 \;
(-1)$. Thus the physical meaning of the binary sequence of a hill in
the exit-velocity fractal is that it gives the sequence of relative
amplitudes between the two waves at the bounce points before the
collision time, with digit $1 \; (-1)$ meaning the left wave is
higher (lower). At the collision time, the two waves are the closest
and interact most strongly. For most of the points on the hill
(except a small section in the middle), the two waves separate from
each other after the collision time, and the exiting waves have
opposite relative amplitudes on the two sides of the hill. The
specific relative amplitudes of the two waves are determined by the
sign of $-\dot{\psi}(\infty)$ which has been given above. The left
wave will have higher (lower) amplitude if $-\dot{\psi}(\infty)$ is
positive (negative).

\section{The map for negative $\varepsilon$ and its predictions
for the PDE/ODE system} \label{nepsi}

\begin{figure}
\includegraphics[width=130mm,height=110mm]{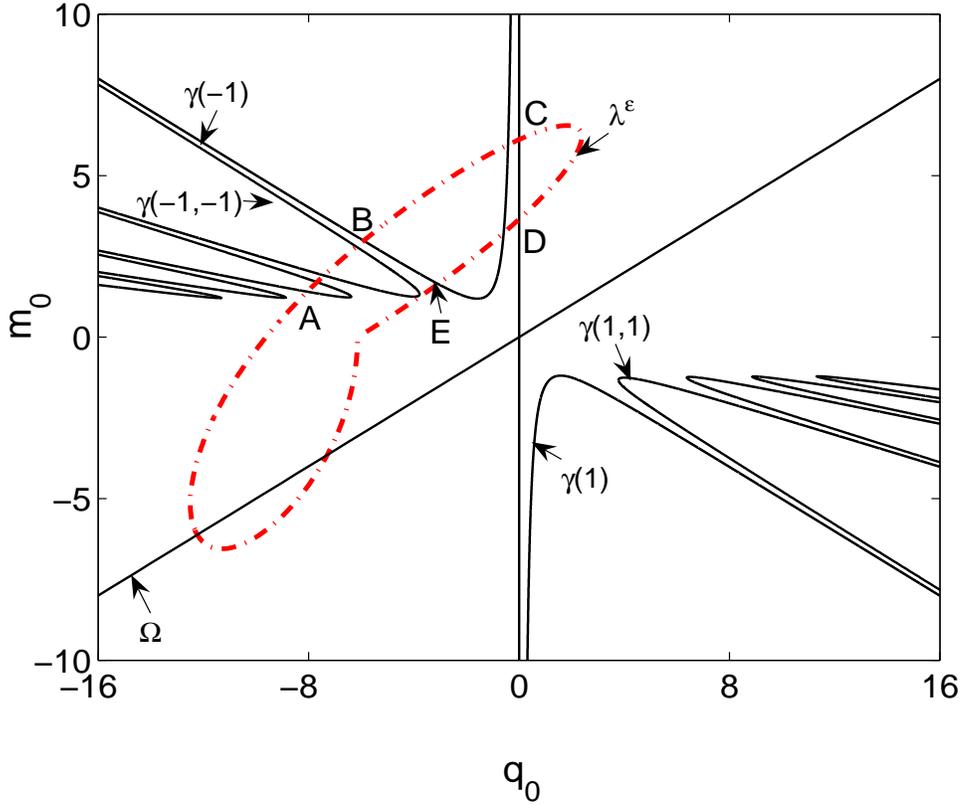}
\caption{\label{nmapsingular} Singular curves of the map
(\ref{pmap1})-(\ref{pmap2}) for negative $\varepsilon$,  as well as
the initial-value curve $\lambda^{\varepsilon}$ for
unequal-amplitude initial conditions (\ref{ic1b}) with
$\varepsilon=-0.001$.  Five of the intersections of
$\lambda^{\varepsilon}$ with singular curves are marked by letters
`A, B, C, D, E'. }
\end{figure}

In this section, we consider the $\varepsilon<0$ case. For the
cubic-quintic nonlinearity (\ref{cqn}), negative $\varepsilon$
occurs when $\delta<0$ \cite{zhuyang}. In this case, we will show
below that fractal scatterings can not occur. The map
(\ref{pmap1})-(\ref{pmap2}) for negative $\varepsilon$ differs from
that for positive $\varepsilon$ by only a sign, but this makes a
crucial difference. Now the map is
\begin{eqnarray}\label{negF}
\mathcal{F}\left(\begin{array}{l}q\\m\end{array}\right)=\left(\begin{array}{l}q+2m+\frac{2
{sgn}(q)}{q^2}\\m+\frac{{sgn}(q)}{q^2}\end{array}\right),
\end{eqnarray}
and
\begin{eqnarray} \label{negFinv}
\mathcal{F}^{-1}\left(\begin{array}{l}q\\m\end{array}\right)=
\left(\begin{array}{l}q-2m\\m-\frac{
{sgn}(q-2m)}{(q-2m)^2}\end{array}\right).
\end{eqnarray}
Notice that $\mathcal{F}$ and $\mathcal{F}^{-1}$ are still
area-preserving and orientation preserving, but the singular-curve
structure changes drastically. To see that, we start with the
singular curves $\gamma(1)$ and $\gamma(-1)$, which are now
\begin{eqnarray}
&&\gamma(1)=\{(q_0,m_0):m_0=-q_0/2-1/q_0^2, \; q_0>0\}, \\
&&\gamma(-1)=\{(q_0,m_0):m_0=-q_0/2+1/q_0^2, \; q_0<0\}.
\end{eqnarray}
These two curves are shown in Fig. \ref{nmapsingular}. We see that
these curves are in the second and fourth quadrants. More
importantly, they do not intersect with $\Omega$. Because of this,
each of these two singular curves has only one pre-image curve under
the map of $\mathcal{F}^{-1}$. This contrasts with the
$\varepsilon>0$ case where each singular curve has two pre-image
curves. The pre-image curves of $\gamma(1)$ and $\gamma(-1)$,
\begin{equation}
\gamma(1,1)= \mathcal{F}^{-1}(\gamma(1)), \quad \gamma(-1,-1)=
\mathcal{F}^{-1}(\gamma(-1)),
\end{equation}
are also shown in Fig. \ref{nmapsingular}. They do not intersect
with $\Omega$ either, thus have only one pre-image curve each as
well. Repeating this process, then all the singular curves one can
get, in addition to the vertical axis $\gamma_0$, are only two
sequences
\[\{\gamma(-1), \; \gamma(-1,-1), \; \gamma(-1,-1,-1), \; ...\} \]
and
\[\{\gamma(1), \; \gamma(1,1), \; \gamma(1,1,1), \;  ...\}, \]
which are shown in Fig. \ref{nmapsingular}. Obviously these two
sequences can not develop fractals. If we draw the initial-value
curves (\ref{case1m}) and (\ref{case2m}) in this same plane, these
curves will intersect with only a finite number of singular curves.
An example is shown in Fig. \ref{nmapsingular}, where the
unequal-amplitude initial conditions (\ref{ic1b}) and
$\varepsilon=-0.001$ are taken. Thus the corresponding exit-velocity
graph in the $\psi_0$ space will have a limited number of
singularity peaks, which is precisely what we observed in Fig.
11(5)-(6) of Ref. \cite{zhuyang}.

\begin{figure}
\begin{center}
\includegraphics[width=100mm,height=90mm]{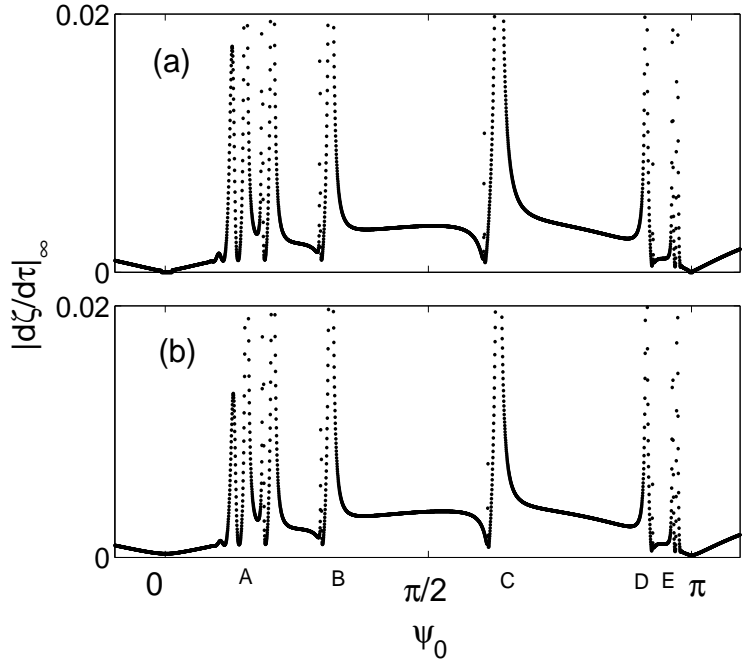}
\caption{\label{necompare} Comparison of the exit-velocity graphs
from (a) ODE predictions, and (b) map predictions for
$\varepsilon=-0.001$. The initial conditions of the ODEs are
(\ref{ic1b}) for unequal-amplitude initial waves. Letters `A, B,
...' are locations of some singularity peaks in (b), which
correspond to the intersections of $\lambda^{\varepsilon}$ with
singular curves in Fig. \ref{nmapsingular} (marked by the same
letters). }
\end{center}
\end{figure}

Now we quantitatively compare the map's predictions with the ODE
results for negative values of $\varepsilon$ (comparison with the
PDE results is expected to be similar, see Figs. \ref{comparison}
and \ref{comparison2}). For this purpose, we take
$\varepsilon=-0.001$ and the unequal-amplitude initial conditions
(\ref{ic1b}) for the ODEs (\ref{eqDyfinal}). In the PDE
(\ref{eqGNLS}) with cubic-quintic nonlinearity (\ref{cqn}), these
ODE initial conditions correspond to $\delta=-0.0003$ and
unequal-amplitude initial conditions (\ref{ic2}) \cite{zhuyang}.
From the direct simulations of these ODEs, the graph of
exit-velocity $|\dot{\zeta}|_\infty$ versus the initial-phase
difference $\psi_0$ is shown in Fig. \ref{necompare}(a). A finite
number of singularity peaks can be seen, and this graph does not
have a fractal structure. To obtain the map's predictions, we use
the initial-value curve (\ref{case2m}) for this case as shown in
Fig. \ref{nmapsingular}. For each point on this initial-value curve,
we iterate the map (\ref{pmap1})-(\ref{pmap2}) to infinity to obtain
the exit velocity from Eq. (\ref{zetaM}). The results of the map's
predictions are shown in Fig. \ref{necompare}(b). As can be seen,
the map's prediction agrees with the ODE results very well. To
better understand these graphs, we label five representative
singularity peaks by letters `A, B, C, D, E' in Fig.
\ref{necompare}(b). Their corresponding points on the initial-value
curve $\lambda^\varepsilon$ of Fig. \ref{nmapsingular} are also
labeled by the same letters. This connection makes it very easy to
understand the exit-velocity graphs of Fig. \ref{necompare}. In
particular, since $\lambda^\varepsilon$ has 10 intersections with
singular curves in Fig. \ref{nmapsingular}, this explains why the
exit-velocity graphs in Fig. \ref{necompare} have 10 singularity
peaks as well. One can further predict the dynamics of the ODE
solution at each $\psi_0$ value in the exit-velocity graph of Fig.
\ref{necompare}(a) in the same way as we did in the previous
section. For instance, if $\psi_0$ is on a hill of binary sequence
$\mathbf{a}=(-1, -1, -1)$ in the exit-velocity fractal of Fig.
\ref{necompare}(a), then the $\zeta(t)$ solution will oscillate
three times before the collision time. The $-\dot{\psi}$ solution
will exhibit spikes at the three bounce points whose sign sequence
is $\mathbf{a}$. After the collision time, the $\zeta(t)$ solution
will go to $-\infty$, while the $\dot{\psi}(t)$ solution will
approach a constant whose sign depends on which side of the hill the
$\psi_0$ value is on. With this knowledge on the ODE solution, we
can then predict the dynamics of the PDE solution, again in the same
way as what we did in the previous section. If the two waves
initially have equal amplitude, where the initial conditions for the
ODEs are (\ref{ic1a}), then the initial-value curve
$\lambda^\varepsilon$ in (\ref{case1m}) will only intersect one
singular curve $\gamma_0$ (i.e. the vertical axis). In this case,
the exit-velocity graph of the ODE will have a unique singularity
point at $\psi_0=0$ and be smooth elsewhere. The corresponding
dynamics of ODE and PDE solutions can be predicted in the same way
as above. With these results, an intimate knowledge is then obtained
for the wave interactions in the $\varepsilon<0$ case.

\section{Conclusion and discussion}
In this paper, we analyzed the separatrix map which governs weak
interactions of solitary waves in the generalized NLS equations. We
showed that when $\varepsilon>0$, this map exhibits a fractal
structure which we delineated by tracking its singular curves. Then
through this map's fractal as well as connections between the map
and the ODEs/PDEs, we reached a very deep understanding on the
fractal structures as well as their interaction dynamics in the ODEs
and PDEs. In addition, we analytically determined how the ODE and
PDE fractals change as the parameter $\varepsilon$ varies, and
showed that these predictions agree well with the numerical results.
Furthermore, we proved a claim made earlier in \cite{zhuyang} that
fractal structures in the ODEs/PDEs for $\varepsilon>0$ bifurcate
from the singularity points in the integrable ODEs. When
$\varepsilon<0$, we showed that the separatrix map does not possess
a fractal structure, hence fractal scatterings can not occur in the
corresponding ODEs and PDEs.

Gathering all our results, we can now give a precise criterion for
the existence of fractal scatterings in weak interactions of
solitary waves in the generalized nonlinear Schr\"odinger equations.
When $|\varepsilon|\ll 1$, fractal scatterings will occur if and
only if the following two conditions are met:
\begin{enumerate}
\item $\varepsilon>0$;
\item The map's initial-condition curve $\lambda^\varepsilon$ [see
(\ref{case1m}) and (\ref{case2m}) for instance], under the
restriction of $H<0$, intersects the map's fractal region (see Figs.
\ref{2dmap}, \ref{iccurve1} and \ref{iccurve2}).
\end{enumerate}
When this criterion is met (i.e. fractal scatterings occur), we can
accurately predict the fractal structure as well as the interaction
dynamics in the exit-velocity graph by simply drawing the
initial-condition curve $\lambda^\varepsilon$ on the map's
$|m_\infty|$ fractal of Fig. \ref{2dmap}. Thus, by now we have
provided a simple answer to a complicated fractal-scattering problem
in weak wave interactions.

The only discrepancy between the PDE's exit-velocity graph and our
ODE/map predictions is at the middle part of each hill, where the
ODE's and map's graphs exhibit peaks of infinite height, but the
PDE's graph dips down instead (see Figs. \ref{comparison} and
\ref{comparison2}). In those parameter regions, the two waves come
together and interact strongly, which makes our reduced ODE model
invalid. We have performed preliminary numerical investigations of
the PDEs at the middle part of each hill in the exit-velocity graph,
and found that the dips in those regions are not smooth. Inside each
dip, we found finer structures which are also fractal-like! These
fractal-like structures inside each dip are apparently the product
of strong wave interactions, thus should be related to fractal
scatterings in solitary wave collisions as reported in
\cite{Campbell1,anninos,Kivshar1,YangTan}. Detailed investigations
of finer structures inside dips of the PDE's exit-velocity graph lie
outside the scope of this paper, and will be left for future
studies.

\section*{Acknowledgments} This work was supported in part by the
Air Force Office of Scientific Research.

\end{document}